\documentclass[aps,prb,twocolumn,showpacs, nobibnotes, nofootinbib, superscriptaddress]{revtex4-1}  

\usepackage{graphicx}  
\usepackage{dcolumn}   
\usepackage{bm}        
\usepackage{amssymb}   

\usepackage{verbatim}
\usepackage[linkcolor=blue,colorlinks=true,citecolor=magenta,filecolor=magenta]{hyperref}  

\usepackage{amsmath,amssymb,amsfonts,mathrsfs, mathtools}  

\DeclarePairedDelimiter\bra{\langle}{\rvert}
\DeclarePairedDelimiter\ket{\lvert}{\rangle}
\DeclarePairedDelimiterX\braket[2]{\langle}{\rangle}{#1 \delimsize\vert #2}

\makeatletter
\renewcommand*{\eqref}[1]{%
  \hyperref[{#1}]{\textup{\tagform@{\ref*{#1}}}}%
}

\newcommand*{\rom}[1]{\expandafter\@slowromancap\romannumeral #1@}

\def\@fnsymbol#1{\ensuremath{\ifcase#1\or *\or \dagger\or \ddagger\or
   \mathsection\or \|\or **\or \dagger\dagger
   \or \ddagger\ddagger \else\@ctrerr\fi}}

\makeatother

\usepackage[shortlabels]{enumitem}

\usepackage{stackrel} 

\usepackage{dsfont} 

\begin{document}


\title{Linear rotor in an ideal Bose gas near the threshold for binding}

\author{Tibor Dome}
\email{td448@cam.ac.uk}
\affiliation{ETH Zurich, Department of Physics, Wolfgang-Pauli-Strasse 27, 8093 Z\"urich, Switzerland}
\affiliation{Institute of Science and Technology Austria (ISTA), Am Campus 1, 3400 Klosterneuburg, Austria}
\affiliation{Institute of Astronomy, University of Cambridge, Madingley Road, Cambridge, CB3 0HA, UK}
\author{Artem G. Volosniev}%
\email{artem.volosniev@ist.ac.at}
\affiliation{Institute of Science and Technology Austria (ISTA), Am Campus 1, 3400 Klosterneuburg, Austria}
\author{Areg Ghazaryan}%
\affiliation{Institute of Science and Technology Austria (ISTA), Am Campus 1, 3400 Klosterneuburg, Austria}
\author{Laleh Safari}
\affiliation{Institute of Science and Technology Austria (ISTA), Am Campus 1, 3400 Klosterneuburg, Austria}
\author{Richard Schmidt}%
\affiliation{Institute for Theoretical Physics, Heidelberg University, 69120 Heidelberg, Germany}
\author{Mikhail Lemeshko}%
\affiliation{Institute of Science and Technology Austria (ISTA), Am Campus 1, 3400 Klosterneuburg, Austria}
\date{\today}

\begin{abstract}
We study a linear rotor in a bosonic  bath within the angulon formalism. Our focus is on systems where isotropic or anisotropic impurity-boson interactions support a shallow bound state. To study the fate of the angulon in the vicinity of bound-state formation, we formulate a beyond-linear-coupling angulon Hamiltonian. First, we use it to study attractive, spherically symmetric impurity-boson interactions for which the linear rotor can be mapped onto a static impurity. The well-known polaron formalism provides an adequate description in this limit. Second, we consider anisotropic potentials, and show that the presence of a shallow bound state with pronounced anisotropic character leads to a many-body instability that washes out the angulon dynamics.
\end{abstract}

\pacs{}
\maketitle

\section{Introduction}

A rotating impurity in a bosonic environment (such as a molecule in superfluid helium-4) can be conveniently studied using the angulon concept~\citealp{lemeshko_schmidt_2015}, which is motivated by the well-known polaron quasiparticle~\citealp{pekar_1946,landau_pekar,froehlich_54,HOLSTEIN1959,HOLSTEIN1959_2}. The angulon has been used in the studies of collective states of many-particle quantum systems\cite{lemeshko_schmidt_2016}, the renormalization of the rotational constant $B$~\cite{lemeshko_2017} and the angular self-localization~\cite{li_2017}.
However, work is still needed to bring our understanding of the angulon to that of the polaron~\citealp{alexandrov,Devreese_2016, Chevy2010review, Massignan2014review,Schmidt2018Review,Scazza_2022}. 

One unresolved question is how angulons form when the underlying rotor-bath interaction supports a bound state. This scenario has not been studied so far for the angulon problem, while it has gained much attention in the Bose polaron problem where it leads to a Bose polaron--molecule crossover~\cite{Rath_2013,Li_2014,Grusdt_2017,Isaule2021RG}. It appears necessary to explore the role of bound states in the angulon formalism because it is known that helium atoms can bind to molecules.
Bound states between  a heavy molecule (e.g., Cl$_2$ or OCS) and helium atoms may be deep and have binding energies a few times larger than the separation energy of a helium atom~\citealp{Barnett1994,toennies_peter}; bound states between a light molecule (such as H$_2$ and D$_2$) and a helium atom are expected to be shallow~\cite{Barnett1994}.
It might be particularly interesting to understand the angulon instability~\citealp{lemeshko_schmidt_2016,Cherepanov_Lemeshko_2017}, a resonant transfer of angular momentum from the impurity to the environment, in the presence of shallow bound states.  Indeed, resonant interactions that typically occur at the threshold for binding could enhance the exchange of angular momentum.\par

To study the angulon in the presence of bound states, we must go beyond the paradigm of past theoretical works on the angulon that were performed within the framework of a linear-coupling microscopic theory; higher-order processes are needed for our study. To formulate a suitable minimal model, we rely on the Bogoliubov approximation, and consider beyond-linear-coupling terms similar to those in the strong-coupling Bose polaron problem~\cite{Rath_2013}.  
To study the resulting Hamiltonian, its ground and excited states, we use exact diagonalization (ED).

As anticipated, we find that the presence of a shallow bound state can lead to a strong momentum exchange, washing out quasiparticle features. This strongly modifies properties of the angulon. However, we also find that the momentum exchange is contingent on a sufficient anisotropic component of the microscopic molecule-boson interaction potential.\par 

For molecules in helium clusters, the molecule-boson interactions are indeed anisotropic~\cite{Pack1984,Atkins1996}, providing a microscopic mechanism for the renormalization of the moment of inertia~\citealp{Kwon2000}. However, the main contribution to the binding energy typically comes from the isotropic part of the potential~\cite{Barnett1993,McMahon1995} and spatial correlations of large clusters are isotropic~\cite{Barnett1993,Dorte1996,McMahon1995}. The situation appears to be similar also for light molecules (such as H$_2$ and D$_2$) whose rotational constant is weakly renormalized in a helium droplet~\cite{Qiang2022Femtosecond}.
Further work is thus needed to find the best experimental conditions for the observation of our findings.
\par

Our paper is organized as follows. In Sec. \ref{sec:theory}, we lay out the theoretical framework for our analysis: the many-body description of an immobile molecular impurity immersed in a Bose-Einstein condensate (BEC). We study the single-boson problem in Sec.~\ref{sec:one_boson}. In Sec.~\ref{sec:s_wave}, we investigate the limit of isotropic boson-molecule interactions, which allows us to connect the limiting cases for the angulon and polaron quasiparticles. Anisotropic interactions are considered in Sec.~\ref{sec:p_wave}, where we also illustrate the ``quantum carousel phenomenon'' for a rotor-boson bound state. In Sec.~\ref{sec:s_and_p}, we investigate the transition from purely isotropic to anisotropic interactions. We conclude in Sec. \ref{sec:concl}. We compare one- and two-phonon results in Appendix~\ref{sec:chevy_vs_non_chevy}, and provide technical details in Appendixes~\ref{sec:ham_deriv}, \ref{sec:am_representation}, \ref{sec:am_algebra} and \ref{sec:ed_details}.

\section{Molecular Impurity in Beyond-Linear-Coupling Regime}
\label{sec:theory}

\subsection{Hamiltonian}
{\it Units.} We use a system of units in which $\hbar \equiv 1$. We shall present distances in units of $(m_{\text{b}}B)^{-1/2}$, where $m_{\text{b}}$ is the mass of a boson and $B$ is the rotational constant. Energies shall be measured in dimensionless form as $E/B$ or $E/[k_n^2/(2m_{\mathrm{b}})]$ with $k_n=(6\pi ^2n_0)^{1/3}$ where $n_0$ is the condensate density; the dimensionless density is $n_0(m_{\text{b}}B)^{-3/2}$. To provide some intuition behind dimensionless numbers, we note that for liquid helium $n_0\simeq 22/ \mathrm{nm}^{3}$ (see, e.g., Ref.~\onlinecite{Kalos1981}); $B$ depends strongly on the the mass of the molecule (for a heavy OCS molecule $B\simeq 0.2$ cm$^{-1}$, and light molecules have $B>1$ cm$^{-1}$). This implies that the dimensionless parameter $n_0(m_{\text{b}}B)^{-3/2}$ is typically of the order of one in current experiments. For some numerical illustrations we shall use $n_0(m_{\text{b}}B)^{-3/2}=1$, which corresponds to $B\simeq 0.63$~cm$^{-1}$.
\par 

{\it Laboratory frame.} We consider a single molecule, modeled as a linear rotor, immersed in a homogeneous Bose-Einstein condensate. In the laboratory frame, the microscopic Hamiltonian that describes this system is conveniently  written as~\cite{lemeshko_schmidt_2015} \begin{equation}\hat{\mathcal{H}}^{\text{lab. fr.}} = \hat{H}^{\text{mol}}+\hat{H}^{\text{bos}}+\hat{H}^{\text{mb}},
\end{equation}
where the three terms describe the molecule, the BEC, and their interaction, respectively.
These terms read as
\begin{align}
\hat{H}^{\text{mol}} = & \ \alpha B\hat{\mathbf{J}}^2, \nonumber \\
\hat{H}^{\text{bos}} = & \ \sum_{\mathbf{k}}\epsilon(k)\hat{a}_{\mathbf{k}}^{\dagger}\hat{a}_{\mathbf{k}} + \frac{1}{2}\sum_{\mathbf{k}, \mathbf{k}', \mathbf{q}}V_{\text{bb}}(\mathbf{q})\hat{a}_{\mathbf{k}'-\mathbf{q}}^{\dagger}\hat{a}_{\mathbf{k}+\mathbf{q}}^{\dagger}\hat{a}_{\mathbf{k}'}\hat{a}_{\mathbf{k}}, \nonumber \\
\hat{H}^{\text{mb}} = & \ \sum_{\mathbf{k}, \mathbf{q}}\hat{V}_{\text{mb}}(\mathbf{q}, \hat{\phi}, \hat{\theta})\hat{\rho}(\mathbf{q})\hat{a}^{\dagger}_{\mathbf{k}+\mathbf{q}}\hat{a}_{\mathbf{k}}.
\label{Ham_l_frame}
\end{align}
Here, we have adopted the formalism of second quantization with bosonic creation, $\hat a^\dagger$, and annihilation, $\hat a$, operators. The kinetic energy of a free boson reads as $\epsilon(k)=k^2/(2m_{\text{b}})$. The molecular angular momentum operator in the laboratory frame is denoted as $\hat{\mathbf{J}}$. The potentials $V_{\text{bb}}$ and $\hat{V}_{\text{mb}}$ describe the boson-boson and molecule-boson interaction, respectively. While $\mathbf{k}, \mathbf{k}'$ and $\mathbf{q}$ are three-dimensional momenta $(\hat{\phi}, \hat{\theta})$ denote the two extrinsic Euler angle operators needed to describe the orientation of a linear rotor. These operators measure the orientation of the molecular impurity: $\hat{\phi}\ket{\phi, \theta}=\phi\ket{\phi, \theta}$, where $\ket{\phi, \theta}$ is an eigenstate of the angular degrees of freedom. The impurity frozen at $\mathbf{r}=0$ has a constant Fourier-transformed Dirac $\delta^{(3)}$ density, $\hat{\rho}(\mathbf{q})=1$.  We have introduced for convenience the dimensionless parameter $\alpha \in [0,1]$ to control the strength of the coupling between different angular momentum states. If $\alpha=0$, there is no energy scale associated with rotation; the orientation of the molecule is fixed. The molecule is free to rotate if $\alpha=1$.\par 

{\it Molecular frame.} To simplify the angular momentum algebra, we work in the molecular frame, where the microscopic interaction potential can be written in real space as a function (and not an operator):
\begin{equation}
V_{\text{mb}}(\mathbf{r})=\sum_{\lambda}V_{\lambda}(r)Y_{\lambda 0}(\theta_r, \phi_r).
\label{interaction_pot}
\end{equation}
Here, $Y_{\lambda \mu}$ are spherical harmonics, and $V_{\lambda}$ is the potential that corresponds to the angular momentum channel $\lambda$. Note that we adhere to the Condon-Shortley phase convention
in the definition of the spherical harmonics~\cite{condon_1970}.\par

The Hamiltonian in the molecular frame, \mbox{$\hat{\mathcal{H}}^{\text{m. fr.}}=\hat{S}^{-1}\hat{\mathcal{H}}^{\text{lab. fr.}}\hat{S}$}, reads as
\small
\begin{align}
\begin{aligned}
\hat{S}^{-1}\hat{H}^{\text{mol}}\hat{S} = &\alpha B(\hat{\mathbf{J}}'-\hat{\text{\boldmath$\Lambda$}})^2, \\
\hat{S}^{-1}\hat{H}^{\text{bos}}\hat{S} = &\sum_{\mathclap{k\neq 0,\lambda \mu}}\omega(k)\hat{b}_{k\lambda\mu}^{\dagger} \hat{b}_{k\lambda\mu}, \\
\hat{S}^{-1}\hat{H}^{\text{mb}}\hat{S} = 
&\sum_{\mathclap{k\neq 0, \lambda}}U_{\lambda}(k)\left[\hat{b}_{k\lambda 0}^{\dagger} + \text{H.c.}\right] \\
&+\sum_{\mathclap{\substack{k, q \neq 0 \\ \lambda l l' \nu}}} {}_1W^l_{l' \lambda}(k,q)C^{l \nu}_{l'\nu \lambda 0}\hat{b}_{kl\nu}^{\dagger} \hat{b}_{ql'\nu} \\ 
&+\sum_{\mathclap{\substack{k, q \neq 0 \\  \lambda l l' \nu}}} {}_2W^l_{l' \lambda}(q,k)C^{l \nu}_{l'\nu \lambda 0}\left[\hat{b}_{ql\nu}^{\dagger} \hat{b}^{\dagger}_{kl'-\nu}+\text{H.c.}\right],
\end{aligned}
\label{full_mol}
\end{align}
\normalsize
where 
$\hat{\text{\boldmath$\Lambda$}}$ is the momentum of the bath, $\hat{\mathbf{J}}'$ is the molecular-frame angular momentum operator, and the sums go over the indices that parametrize the angular-momentum basis (see Appendix~\ref{sec:ham_deriv} for technical details). Note that the bosonic creation and annihilation operators were renamed from $\hat{a}$ to $\hat{b}$ to emphasize that the Bogoliubov approximation~\cite{Bogolyubov_1947} was employed in the derivation of Eq.~(\ref{full_mol}). We also transitioned from the single-particle momentum basis defined by $\mathbf{k}$ to the angular momentum basis defined by $(k,\lambda,\mu)$ (see Appendix~\ref{sec:am_representation}). The Bogoliubov quasiparticles are characterized by the standard dispersion relation $\omega(k)$ [see Eq.~\eqref{e_bog_disp}]. $U_\lambda$ determines the strength of the one-phonon coupling, 
\begin{equation}
U_{\lambda}(k) = \sqrt{\frac{2k^2n_0\epsilon(k)}{\omega(k)\pi}}\int_{0}^{\infty}drr^2V_{\lambda}(r)j_{\lambda}(kr),
\label{U_lambd_def}
\end{equation}
whereas the two-phonon coupling coefficients are parametrized as 
\begin{equation}
{}_1W^l_{l' \lambda}(k,q) = (u_ku_q+v_kv_q)\sqrt{\frac{2l+1}{4\pi}}W_{l' \lambda}^l(k,q),
\label{W_1_def}
\end{equation}
and
\begin{equation}
{}_2W^l_{l' \lambda}(k,q) = u_kv_q\sqrt{\frac{2l+1}{4\pi}}W_{l' \lambda}^l(k,q),
\label{W_2_def}
\end{equation}
where $u_k$ and $v_k$ are the Bogoliubov coefficients [see Eq.~\eqref{Bog_solved}] and 
\small
\begin{equation}
W_{l'\lambda}^{l}(k,q)= \frac{2 kq}{\pi}\sqrt{\frac{2l'+1}{2l+1}}C^{l0}_{l'0\lambda 0}\int_{0}^{\infty}dr r^2 V_{\lambda}(r)j_l(kr)j_{l'}(qr).
\label{two_phon_coupling}
\end{equation}
\normalsize
The Clebsch-Gordan coefficients $C^{l\nu}_{l'\nu'\lambda\mu}$ and the spherical Bessel functions of the first kind $j_l(x)$ in the last expression indicate that the rotating impurity couples different angular momentum channels. This is in contrast to a three-dimensional Bose polaron where a single (linear) momentum is sufficient to describe the system (see the
studies at the level of the Fr\"ohlich Hamiltonian~\cite{Tempere_2009,Grusdt_2015,Nielsen2019}), which corresponds to Eq.~(\ref{full_mol}) with $W=0$, and beyond\cite{Rath_2013,Ardila_2015,Christensen_2015,Shchadilova2016, Grusdt_2017, Kain_2018,Hryhorchak_2020,Massignan_2021,Ichmoukhamedov_2019,Pascual_2021,Christianen_2023}.

\subsection{Parameters}
The decomposition of Eq. \eqref{interaction_pot} into spherical harmonics can be used to describe any microscopic interaction potential. However, to understand the qualitative differences between a featureless and a rotating impurity, it is convenient to restrict the angular momentum channels to $\lambda \in \lbrace 0,1\rbrace$. This approximation 
defines a minimal toy model that allows for an exchange of one unit of angular momentum between the molecule and the bath in a single scattering event. Furthermore, from the symmetry considerations, this could be a faithful description for molecules whose point group is $C_{\infty v}$. [The typical examples of this symmetry class such as HF, DF, HCl~\cite{Rodwell1981,Lovejoy1990,Moszynski1994HeHF}, and CO~\cite{Chuaqui1994} might, however, also require $\lambda = 2$ channels.]

Although the approximation $\lambda \in \lbrace 0,1\rbrace$ is natural for our qualitative study~\cite{lemeshko_schmidt_2015}, one might need to include other channels for quantitative description of some molecules in helium droplets. For example, homonuclear molecules (such as Cl$_2$) have  $D_{\infty h}$ symmetries rendering channels with even $\lambda$, especially $\lambda=2$, more prevalent.
\par

For simplicity, we parametrize the multipoles of Eq.~\eqref{interaction_pot}  by attractive Gaussians
\begin{equation}
V_{\lambda}(r)=-\frac{g_{\lambda}}{2r_{\lambda}^2}e^{-r^2/r_{\lambda}^2},
\label{V_model_pot}
\end{equation}
 with interaction strengths $g_0$ and $g_1$. We fix the width of the potential: $r_0=r_1=1.5 \ (m_{\text{b}}B)^{-\frac{1}{2}}$ and vary $|g_0|\sim |g_1| \sim 50 \ m_{\mathrm{b}}B^2$ in our numerical simulations. 
 Even though this model potential does not quantitatively reproduce any particular atom-molecule potential, it allows for the angular momentum exchange, which is the key ingredient of our study. The parameters of the potential are motivated by previous studies of the angulon quasiparticles\cite{lemeshko_schmidt_2015} -- we shall use them to compare and contrast to the already existing results.
\par 

As our focus is on the formation of a bound state in the bath, we set $a_{\text{bb}}=0$ as we do not expect that weak boson-boson interactions can strongly modify this dynamics. In this case, the Bogoliubov rotation itself is trivial, i.e., $u_k \equiv 1, \ v_k \equiv 0$. This eliminates from Eq.~(\ref{full_mol}) the pairing term involving $\hat{b}^{\dagger}\hat{b}^{\dagger}$ and its complex conjugate, which correspond to creation and annihilation of phonon pairs, respectively. We are left with only one beyond-linear term, which involves $\hat{b}^{\dagger}\hat{b}$, and our study represents an investigation of this term within the angulon formalism. This term acts as an external potential for phonons that can support a rotor-boson bound state.

 We note that $a_{\text{bb}}=0$ is the standard limiting case in the study of Bose polarons~\cite{Volosniev2015,Shchadilova2016,Drescher2021,Panochko2021} that is known to describe qualitatively various properties of the system, in particular, its energy~\cite{Rath_2013}. For quantitative calculations, which are beyond this work, finite boson-boson interactions should be taken into account, see, e.g., Refs.~\onlinecite{Rath_2013,Boudjema2014,Shchadilova2016}.

\subsection{Technical Details and Definitions}
\label{ss_terminology}

While we provide an extended summary of the employed technical framework in Apps. \ref{sec:ham_deriv}, \ref{sec:am_representation}, \ref{sec:am_algebra} and \ref{sec:ed_details}, we find it fitting to provide some useful details below. 

{\it Conservation of total angular momentum.} We exploit the fact that the total angular momentum $\hat{\mathbf{L}} = \hat{\mathbf{J}}+\hat{\text{\boldmath$\Lambda$}}$ is conserved to constrain the basis [see the discussion around Eq.~\eqref{composite_am}]. In the absence of an external field that breaks the spherical symmetry, all the relevant physical properties are independent of the quantum number $M$ (associated with $\hat{L}_{Z}$, the projection of the total angular momentum onto the laboratory-frame $Z$ axis), which we will omit hereafter.\par 

{\it Spectral function.} To quantify the response of the system, we work with the zero-momentum spectral density function (see Appendix~\ref{sec:ham_deriv})
\begin{equation}
A(E) = -\frac{1}{\pi}\text{Im} \lim_{\epsilon \rightarrow 0^+} \ \sum_L \sum_{j}\frac{|\braket{\Psi_{\mathrm{NI}}}{\Psi_{L}^{(j)}}|^2}{E-\varepsilon_L^{(j)}+i\epsilon}.
\label{spect_func_def}
\end{equation}
The $j$ index runs over all eigenstates $\Psi_L^{(j)}$ of $\hat{\mathcal{H}}^{\text{m. fr.}}$ for a given total angular momentum $L$. The corresponding eigenvalues are $\varepsilon_L^{(j)}$. The spectral function provides information about the excitations of the system that have a ``significant'' overlap with the non-interacting state $\Psi_{\mathrm{NI}}$ (no bosonic excitations). Even though the spectral function in Eq.~\eqref{spect_func_def} obeys a sum rule (see Appendix~\ref{sec:ham_deriv}), the maximum value of $A(E)$ has no significance for our study. Hence, for plotting purposes we linearly rescale $A(E)$ such that its maximum at any frequency is $1$. We illustrate our results using $\epsilon=0.05 \ B$. \par 

{\it One-phonon vs two-phonon excitation ansatz.} The one-phonon or single-phonon ansatz means that the state vector used in ED includes only single-phonon excitations (see Appendix~\ref{sec:ed_details} for details). It resembles the Ch\'evy ansatz, originally introduced for an imbalanced Fermi gas~\cite{Chevy_2006}.  Accordingly, two phonon or double phonon shall refer to results obtained with up to two-phonon excitations in ED.\par 

{\it Polaron picture vs angulon picture.} To analyze the molecular impurity problem, we present our results in two different ways (``pictures'') depending on which parameters are being varied. We shall refer to the first one as the \textit{polaron picture}, as it is standard for studies of Bose polarons. In this picture, the behavior of the system is analyzed as a function of either the $s$-wave scattering length $a_{\text{ib}}$ (which in cold-atom experiments can be tuned by means of a Feshbach resonance~\cite{Fukuhara_2013, Catani_2012}) or the microscopic interaction strengths $g_i$.\par 

The mapping from the $s$-wave interaction strength $g_0$ to the scattering length $a_{\text{ib}}$ is done using a numerical (RK4) scheme that solves the Schr\"odinger ordinary differential equation (ODE) for $a_{\text{ib}}$~\cite{Jeszenszki_2018}, keeping $r_0 = 1.5 \ (m_{\text{b}}B)^{-\frac{1}{2}}$ fixed. For anisotropic potentials, even though effective range parameters also exist, we prefer to investigate the energetics of the system directly as a function of $g_1$ in terms of the dimensionless parameter $(g_1m_{\mathrm{b}})^{1/4}k_n^{-1}$. 

Note that the standard choice for the dimensionless energy in studies of the Bose polaron, $E/[k_n^2/(2m_{\mathrm{b}})]$, is not natural for presenting the energy spectrum for molecular impurities. In particular, it leads to big gaps between different $L$ blocks. Therefore, to illustrate our findings, we set $\alpha = 0.4$ for calculations reported in the polaron picture (except for Fig. \ref{fig:N1_polaron}, bottom panel).\par 

We shall refer to the second way of presenting results as the \textit{angulon picture}. It is standard for studies of the angulon quasiparticle. In this picture, spectral properties are expressed as a function of the dimensionless bath density $n_0(m_{\mathrm{b}}B)^{-3/2}$ or $\ln\left(n_0(m_{\mathrm{b}}B)^{-3/2}\right)$. This is practical for molecular impurities since in the limit $n_0 \rightarrow 0$, we obtain a free rotor whose rotational constant $B$ is that of the non-interacting molecule.\par

By presenting results in two different pictures, we hope that our results will be easily understood by researchers working on Bose polarons as well as on angulons. As we illustrate in the next sections, the polaron picture allows us to demonstrate the evolution of the energy spectrum close to the threshold for binding in a simple way. At the same time, we find the angulon picture suitable for showing a transition between isotropic and anisotropic interaction potentials.

\section{One-boson problem}
\label{sec:one_boson}

\subsection{Without the bath}
To set the stage for our investigation of a many-body system, we start with the single-boson problem, which corresponds to removing the one-phonon coupling term and setting the bath density $n_0$ to zero in Hamiltonian~\eqref{full_mol}. This leads to a single boson that interacts with the impurity, allowing us to isolate the physics that comes from the beyond-Fr{\"o}hlich term. By toggling between $\alpha = 0$ and $\alpha = 1$, we can have a static (``very heavy'') or a rotating molecule, respectively.\par

\begin{figure}[t]
\hspace{-0.3cm}
\includegraphics[scale=.535]{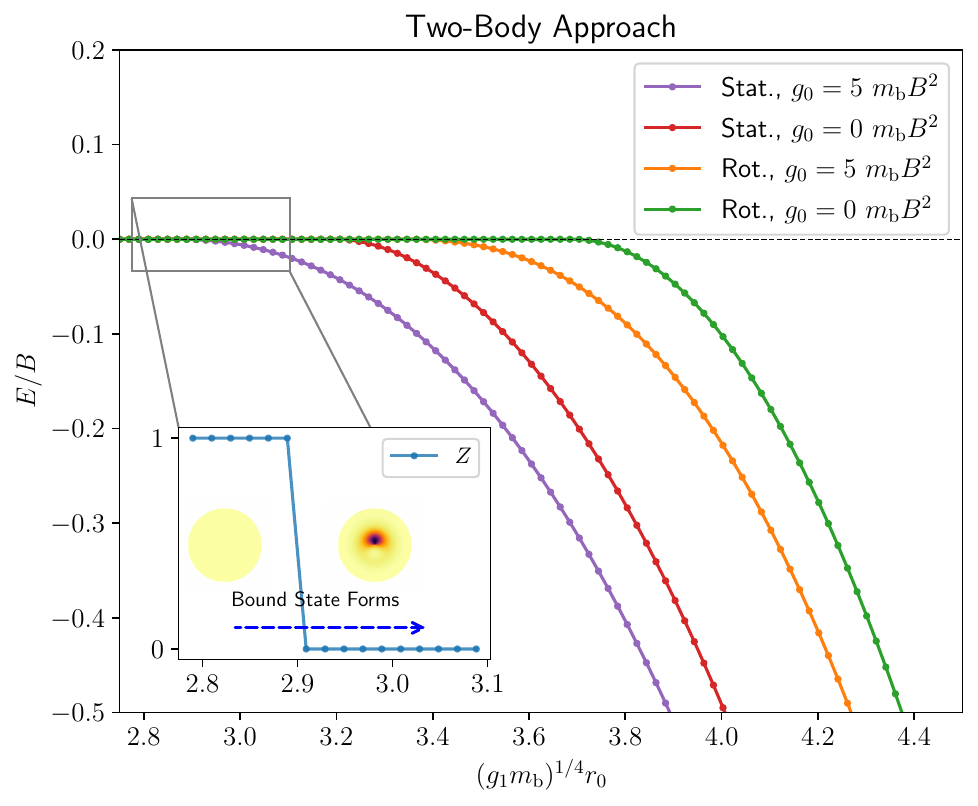}
\caption{Ground-state energy of a boson interacting with a static (purple and red) and rotating (orange and green) molecule for two values of $g_0$. The inset shows the ``quasiparticle weight'' $Z$ for the purple line as a function of the rescaled $g_1$. The bosonic density profiles $n_{\text{ph}}(\mathbf{r})$ calculated using Eq.~\eqref{phon_dens_expr} in the molecular frame are shown for $(g_1m_b)^{1/4}r_0 = 2.8, \ 3.0$.}
\label{fig:bound_states}
\end{figure}

In Fig.~\ref{fig:bound_states}, we show the ground-state energy for the lowest total angular momentum block $L=0$ as a function of $g_1$ for fixed $g_0$. Not surprisingly, we find that a rotating molecule starts to support a bound state  at a higher value of $g_{1,\mathrm{crit}}$ compared to a static molecule. This is a direct consequence of the rotational kinetic energy $B(\hat{\mathbf{J}}'-\hat{\text{\boldmath$\Lambda$}})^2$ that counteracts bound-state formation (cf.~Ref.~\onlinecite{Blinov2004}). There is a resemblance of this result to the standard textbook two-body problem~\cite{Jensen_2004} where the reduced mass increases the overall kinetic energy and changes the threshold for binding. However, a rotating impurity cannot be reduced to a static one with some renormalized parameters, making the problem we consider physically richer and numerically more difficult.\par

For $g_1 < g_{1,\mathrm{crit}}$, the boson is not bound to the impurity, and its 
 probability density [defined in Eq.~\eqref{phon_dens_expr} in the molecular frame] is given by $1/V$, where $V$ is the volume of the system. At $g_1>g_{1,\text{crit}}$, a bound state exists, leading to a finite probability to find the boson in the vicinity of the molecule. This is evident from the density profile in the inset of Fig.~\ref{fig:bound_states}, which features a $p$-wave lobe characteristic for a $p$-wave dominated model potential, Eq.~\eqref{V_model_pot}. The size of the $p$-wave lobe decreases as $g_1$ increases, reflecting a more compact bound state.\par
 
The formation of the bound state can be detected by considering the overlap of the ground state $\Psi_0$ with the non-interacting state $\Psi_{\mathrm{NI}}$:
\begin{equation}
Z=|\braket{\Psi_0}{\Psi_{\mathrm{NI}}}|^2.
\end{equation}
This quantity, which turns into the residue of a quasiparticle for a many-body problem,
 features (for $V\to\infty$) an abrupt change at $g_1=g_{1,\text{crit}}$
from one to zero, which implies a vanishing spectral function in Eq.~\eqref{spect_func_def}. As the attractive Gaussian interaction strength $g_0>0$ is gradually increased in magnitude, the value of $g_{1,\mathrm{crit}}$ decreases, but the character of bound-state formation remains unaltered (abrupt change of $Z$ at $g_1 = g_{1,\mathrm{crit}}$; vanishing ground-state energy, $E=0$, for $g_1 < g_{1,\mathrm{crit}}$).\par 

\subsection{In the presence of the condensate}
\begin{figure}[t]
\hspace{-0.6cm}
\includegraphics[scale=.52]{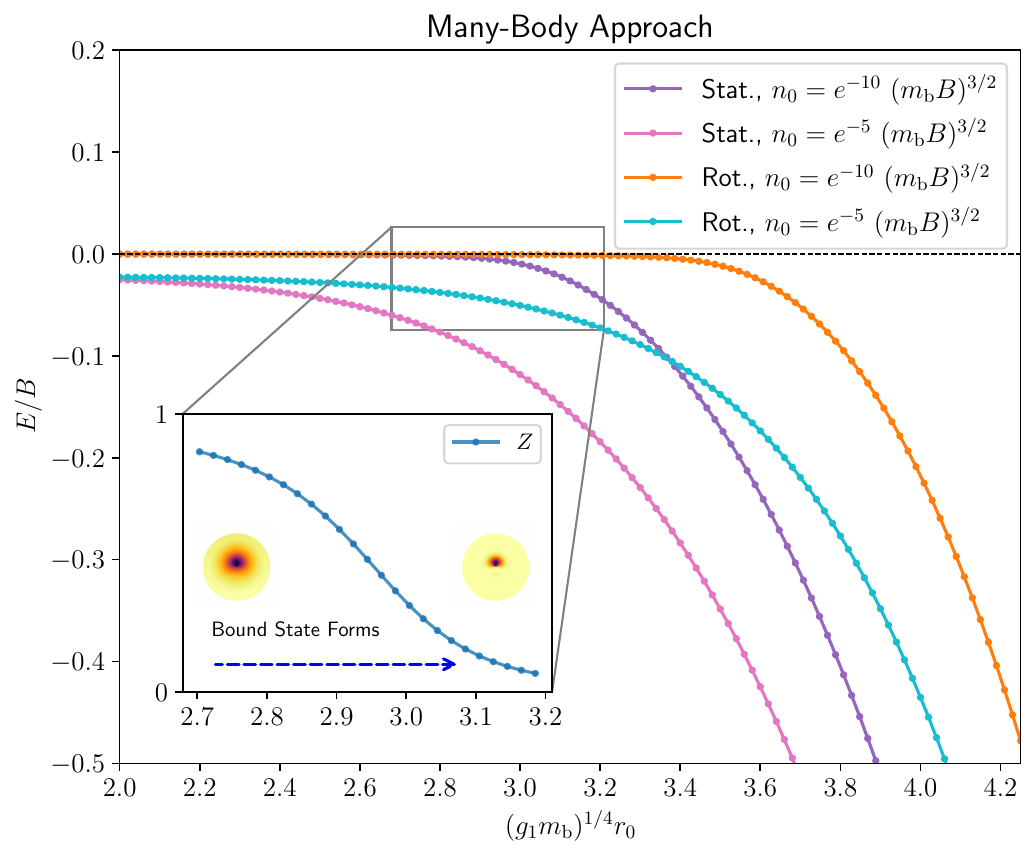}
\caption{Ground-state energy of a boson interacting with a static (purple and pink) and rotating (orange and cyan) molecule for $g_0=5 \ m_{\mathrm{b}}B^2$ in the presence of a condensate of varied density $n_0$. The inset shows the quasiparticle weight $Z$ for the purple line as a function of $g_1$ (compare this inset to the one of  Fig.~\ref{fig:bound_states}). The bosonic density profiles in the molecular frame are shown in the inset on either side of the threshold for binding, at values $(g_1 m_b)^{1/4}r_0 = 2.7, \ 3.2$.}
\label{fig:bound_states_manybody}
\end{figure}

Let us now consider what happens to the one-boson physics discussed above in the presence of the condensate. To this end, we study the many-body Hamiltonian from Eq.~\eqref{full_mol} with a single excitation \textit{on top} of a dilute condensate with a given density. We present the corresponding  ground-state energy in Fig.~\ref{fig:bound_states_manybody}, which illustrates the transition from the physics dominated by the Fr{\"o}hlich-type Hamiltonian to the one driven by bound state formation. 

As shown in Fig.~\ref{fig:bound_states_manybody} for $g_1<g_{1,\text{crit}}$, the linear-coupling terms drastically modify the behavior of the system already for very small bath densities. In particular, the ground-state energy of the system is negative even if the $\hat{b}^{\dagger}\hat{b}$ term does not support a bound state, as e.g. for $g_0=5 \ m_{\mathrm{b}}B^2$, $n_0 = e^{-10} \ (m_{\mathrm{b}}B)^{3/2}$, and small values of $g_1$. For weak couplings the reduction of the energy simply follows from the fact that the second-order correction to the ground-state energy in perturbation theory is always negative. In fact, for the Fr{\"o}hlich-polaron problem (no $b^\dagger b$ term) one finds~\cite{LLP_1953,Devreese_2016}
\begin{equation}
E_0 = -\sum_{\mathbf{k}}\frac{|U_0(k)|^2}{\omega(k) + k^2/(2m_{\mathrm{b}})},
\label{eq:EO_froehlich}
\end{equation}
where $U_0(k)$ is the interaction potential in the zero angular
momentum sector as defined in Eq.~(\ref{U_lambd_def}).
As is evident from Eq.~(\ref{eq:EO_froehlich}), systems with finite densities
 have a negative value of the ground-state energy, $E<0$. This is in contrast to Fig.~\ref{fig:bound_states} where the energy is identically zero for $g_1 < g_{1,\mathrm{crit}}$.\par

Moreover, bound-state formation in the presence of the environment features a smooth decrease of $Z$ to zero. Indeed, even for a small non-zero bath density such as $n_0 = e^{-10} \ (m_{\mathrm{b}}B)^{3/2}$, there is a finite probability to find a boson close to the impurity, which turns the bound-state formation into a gradual process. This is similar to the BEC-induced hybridization between a polaron and molecular state described in the context of strong-coupling Bose polarons~\cite{Rath_2013}. Finally, we note that the critical value of $g_{1,\mathrm{crit}}$ for binding cannot be identified unambiguously for finite densities. However, for practical purposes, one may identify the interaction strength at which bound-state formation occurs with the rule-of-thumb criterion $Z(g_{1,\mathrm{crit}})=0.5$.\par

\section{Isotropic interactions: Static impurity limit}
\label{sec:s_wave}

As a next step, we consider isotropic interactions, i.e., we assume that $g_1=0$.
In this limit, the boson-impurity interaction cannot change the angular momentum of the molecule, meaning that it is a good quantum number and that the angulon is not formed. Still, isotropic interactions provide a useful limit for estimating the energies, and understanding spatial correlations~\cite{Barnett1993,Kwon1996,McMahon1995,Mikosz2006}. 
As different rotational states of the molecule are decoupled, they can be seen as a static limit of the Bose polaron problem: an impurity in a BEC~\citealp{Hu_2016,Jorgensen_2016,Yan_2020} or superfluid helium~\citealp{Toennies_1998}. Therefore, we can adopt technical tools and physical intuition developed for Bose polarons. In particular, for the analysis here, we use the polaron picture (see Sec.~\ref{ss_terminology}).

\begin{figure}
\includegraphics[scale=0.59]{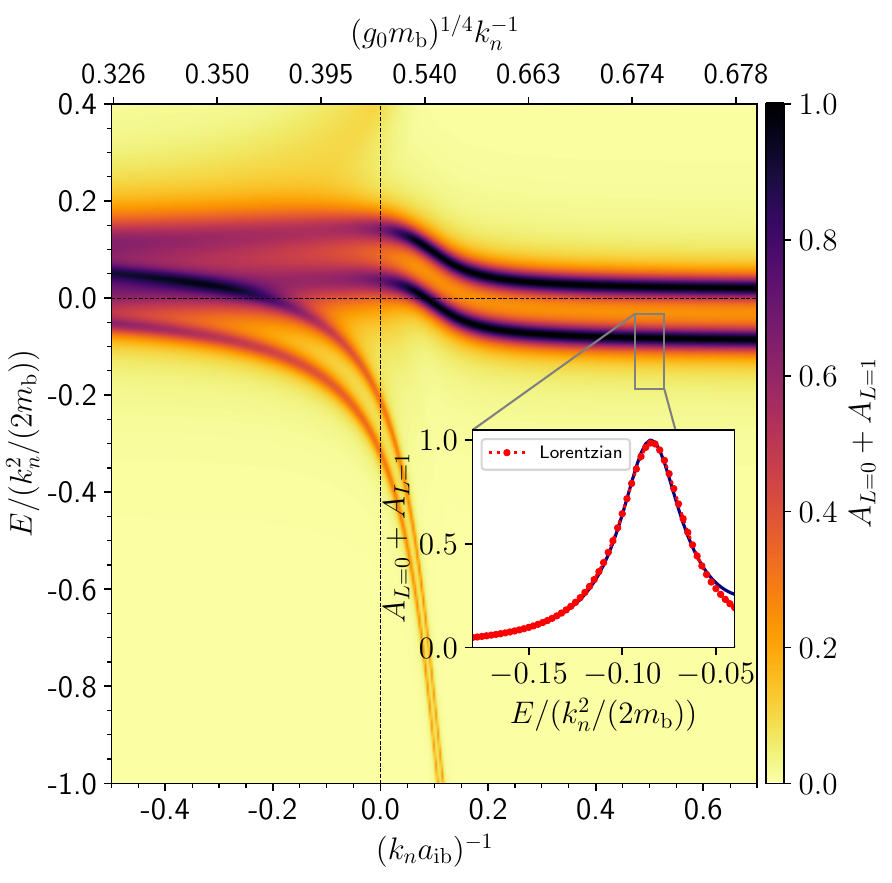}
\caption{\label{fig_lin_vs_quasi_scatt} Spectral function of a system with $g_1=0$ calculated using two-phonon ED for the $L=0,1$ blocks with $\alpha = 0.4$ and bath density $n_0 = 1.0 \ (m_{\mathrm{b}}B)^{3/2}$. The dependence on both $(k_na_{\text{ib}})^{-1}$ and $(g_0m_{\mathrm{b}})^{1/4}k_n^{-1}$ is shown, illustrating the non-linear mapping $a_{\text{ib}}  \leftrightarrow g_0$. The inset demonstrates the spectral function profile at $(k_na_{\text{ib}})^{-1} = 0.5$ (blue solid curve) together with its Lorentzian best fit (red dashed curve).}
\end{figure}

In Fig.~\ref{fig_lin_vs_quasi_scatt}, we show the spectral function~\eqref{spect_func_def}, on two-phonon level for the blocks $L=0,1$ assuming a bath density of $n_0 = 1.0 \ (m_{\mathrm{b}}B)^{3/2}$. As anticipated, we find that the angulon reduces to a ``static polaron'' in each $L$ block. The well-known\cite{Rath_2013,Scazza_2022} attractive and repulsive polaron branches are recovered. Importantly, the polarons in the different $L$ blocks are exact \textit{replicas} of one another, except for a trivial offset in energy due to the rotational kinetic energy, $\alpha B L (L+1)$. In particular, spectral weights in the $L=0$ and $1$ blocks match identically. 


As we are using a Gaussian potential, finite-range effects are visible in the spectral function. In particular, we see that the ``repulsive polaron'' feature can have negative energies. In fact, we understand this feature as not the standard repulsive polaron but as a particle scattering off a molecule-boson bound state. Indeed, this feature appears only at the two-phonon ED level. Furthermore, its spectral function has a well-defined Lorentzian profile (see Fig.~\ref{fig_lin_vs_quasi_scatt}) that hints at the resonant scattering process. We discuss finite-range effects further in Appendix.~\ref{sec:chevy_vs_non_chevy}.\par  

\begin{figure}[t]
\vspace{0.35cm}
\includegraphics[scale=.57,trim={0.2cm 0 0 0},clip]{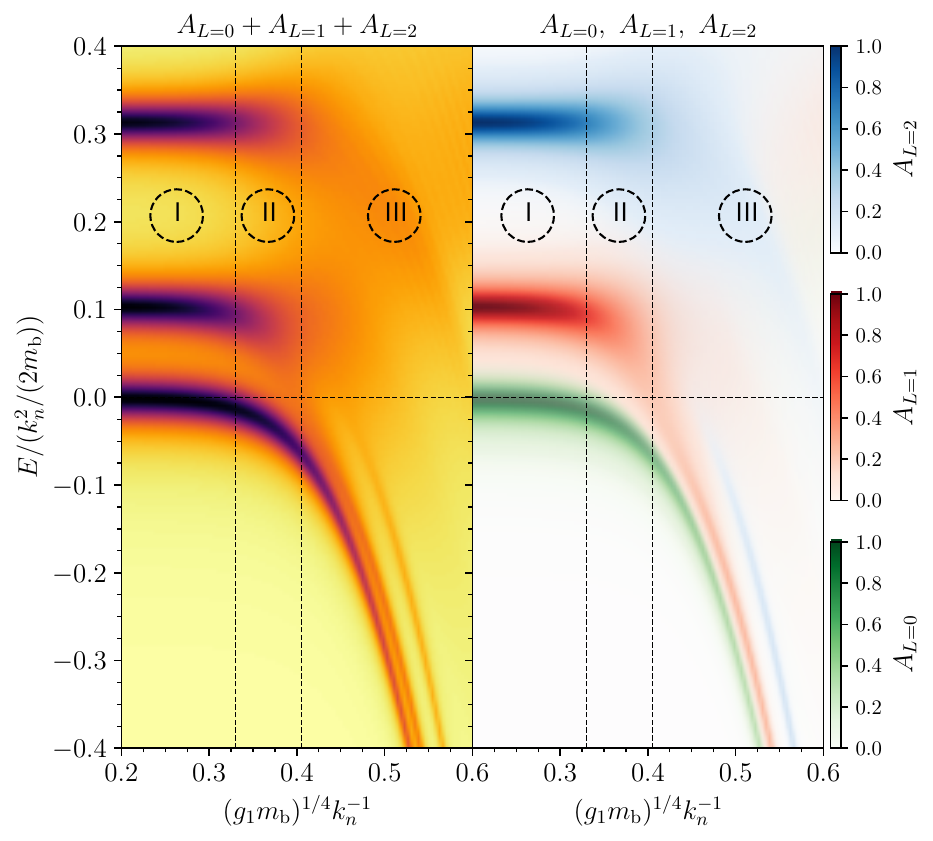}
\caption{Left panel: Spectral function of the $p$-wave dominated ($g_0=0$) angulon calculated using two-phonon ED, for the $L=0, 1, 2$ blocks with $\alpha = 0.4$ and bath density $n_0 = 1.0 \ (m_{\mathrm{b}}B)^{3/2}$. The color map is identical to the one in Fig. \ref{fig:N1_polaron}. Right panel: Decomposition of the spectral function into the three $L$ blocks using color maps.}
\label{fig:hybrid_formalism}
\end{figure}

\section{Anisotropic Interactions}
\label{sec:p_wave}

Having analyzed the case of $g_1 = 0$, let us study the other limiting case of only anisotropic interactions, i.e., $g_0=0$. Such $p$-wave potential dominated systems feature rich spectral characteristics arising from many-body effects that cannot be reduced to polaron physics. To present our results, we adopt both the polaron and angulon pictures.\par

\subsection{Polaron picture}

In Fig. \ref{fig:hybrid_formalism}, we show the  spectral function of the angulon with the total angular momentum up to $L=2$ at a fixed bath density, $n_0 = 1.0 \ (m_{\mathrm{b}}B)^{3/2}$. We plot the spectral function against the dimensionless interaction parameter $(g_1m_{\mathrm{b}})^{1/4}k_n^{-1}$. It is practical to identify three regimes in our data:

\begin{enumerate}[i]
\item \textit{Dressed molecule}\label{enum_i}. In the weak-coupling regime $(g_1m_{\mathrm{b}})^{1/4}k_n^{-1} \lesssim 0.32$, the dressed rotor behaves as a well-defined quasiparticle with large spectral weight. The two-body potential supports no bound states, and the rotor can be visualized as a ``carousel'' with no rider.
\item \label{item_mb_instab}\textit{Many-body instability}\label{enum_ii}. The two-body potential with a critical interaction strength \mbox{$(g_1m_{\mathrm{b}})^{1/4}k_n^{-1} \simeq 0.32$} supports a shallow bound state for $L=0$. This implies a resonant bath-impurity interaction that alters the angulon spectrum drastically. For $L\neq 0$, the immersed molecule enters a superposition of a free rotor (non-interacting state) and phononic states (at most two excitations in Fig.~\ref{fig:hybrid_formalism}) on top of the condensate, which is seen as an apparent discontinuity in the response of the system.\par

We refer to this regime as a many-body instability, which arises due to the resonant transfer of angular momentum between the impurity and the bath via the anisotropic interaction ($g_1 \neq 0$) (cf.~Ref.~\onlinecite{lemeshko_schmidt_2015}). In terms of a quantum carousel analogy, this regime can be visualized as an ``unsuccessful attempt of phonons to jump into a seat of a carousel''. 

\item \textit{Bound states}\label{enum_iii}. For interaction strengths larger than $(g_1m_{\mathrm{b}})^{1/4}k_n^{-1} \simeq 0.41$ we observe formation of one attractive branch (bound state) per $L$ block, decreasing in energy with increasing $g_1$: a quantum `carousel' with a rider. 
\par

The bound-state regime for $L=1$ seems to be realized for stronger couplings in comparison to $L=2$. This observation can again be understood using the `carousel' interpretation in which rotational kinetic energy competes with the attractive force of the impurity. Note that the $L$ blocks come closer in energy in comparison to the $g_1=0$ case. This can be rationalized by noticing that the moment of inertia of the bound state is larger than that of a free molecule. In particular, we observe that for $(g_1m_{\mathrm{b}})^{1/4}k_n^{-1}\gtrsim 0.45$ the spectrum can be approximated by $B_{\mathrm{eff}}L(L+1)$, where $B_{\mathrm{eff}}<B$. This can be interpreted as an angulon where instead of a molecule the basic element is a bound state between an impurity and a boson.
\end{enumerate}

\begin{figure}[t]
\includegraphics[scale=.43,trim={0.25cm 0 0 0},clip]{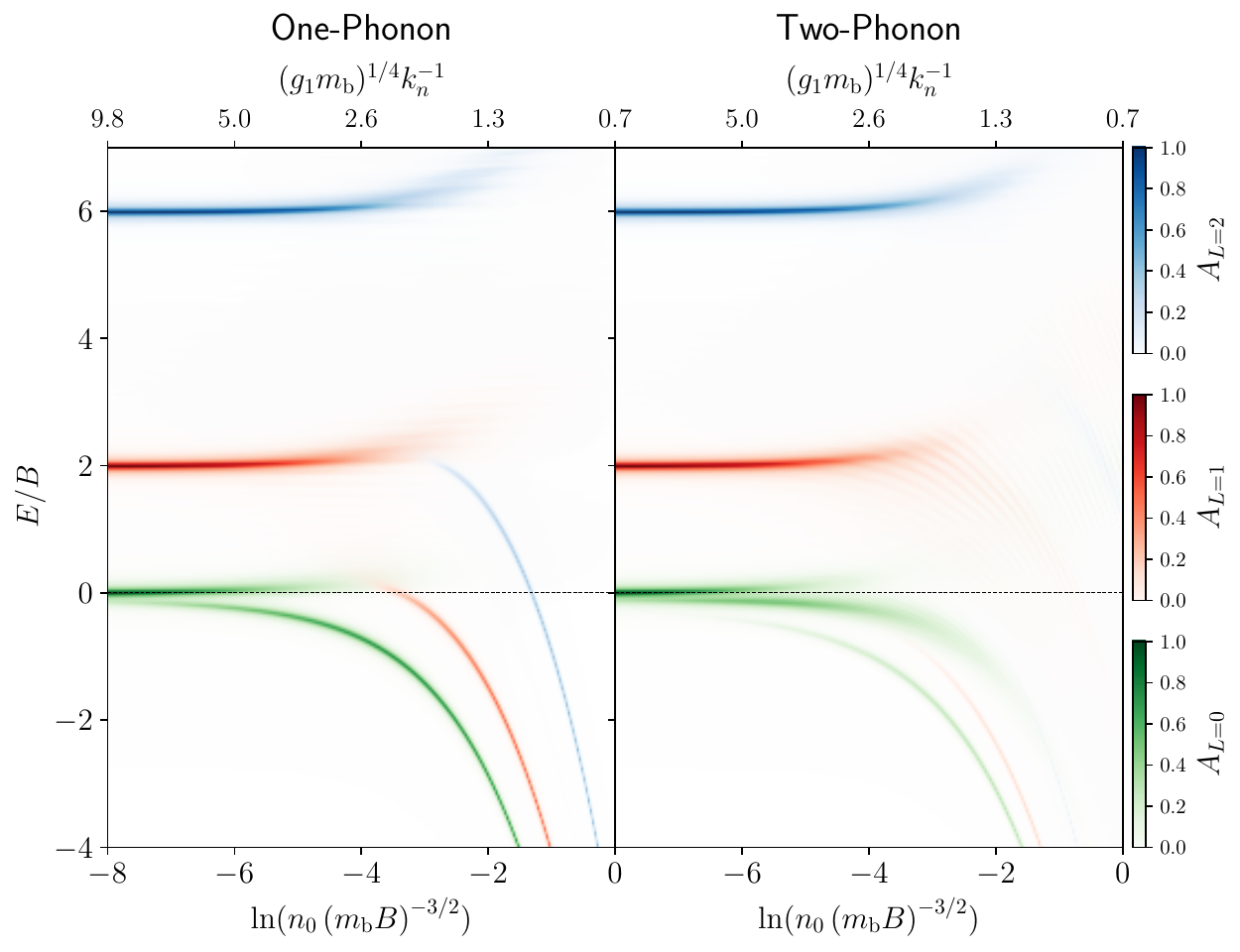}
\caption{Spectral function of the $p$-wave dominated ($g_0=0$) angulon, decomposed into $L=0,1,2$ blocks with $\alpha = 1$. The interaction strength is $g_1=50 \ m_{\mathrm{b}}B^2$, chosen such that for $L=0$ there is one bound state in the limit $n_0\to0$. In the dimensionless units of Fig.~\ref{fig:bound_states_manybody}, the interaction strength is $(g_1m_b)^{1/4}r_0 = 4.0$. To facilitate a comparison with Fig.~\ref{fig:hybrid_formalism}, we have added the dependence on $(g_1m_{\mathrm{b}})^{1/4}k_n^{-1}$ above the figure. The left (right) panel is calculated using one-phonon (two-phonon) ED.}
\label{fig:spec_lin_angulon}
\end{figure}
 
To summarize regimes \eqref{enum_ii} and \eqref{enum_iii}: in the instability regime, the spectral weight of the angulon quasiparticle becomes negligible. Once the interaction strength becomes sufficiently strong, a long-lived angulon state based on a bound state emerges. The formation of the bound state in our ED calculations is similar to the polaron-molecule transition where a particle is taken from the bath to form a molecule~\cite{Grusdt_2017, Ness_2020, Peng_2021}.\par 

\begin{figure*}[t]
\includegraphics[scale=.6,trim={0.2cm 0 0 0},clip]{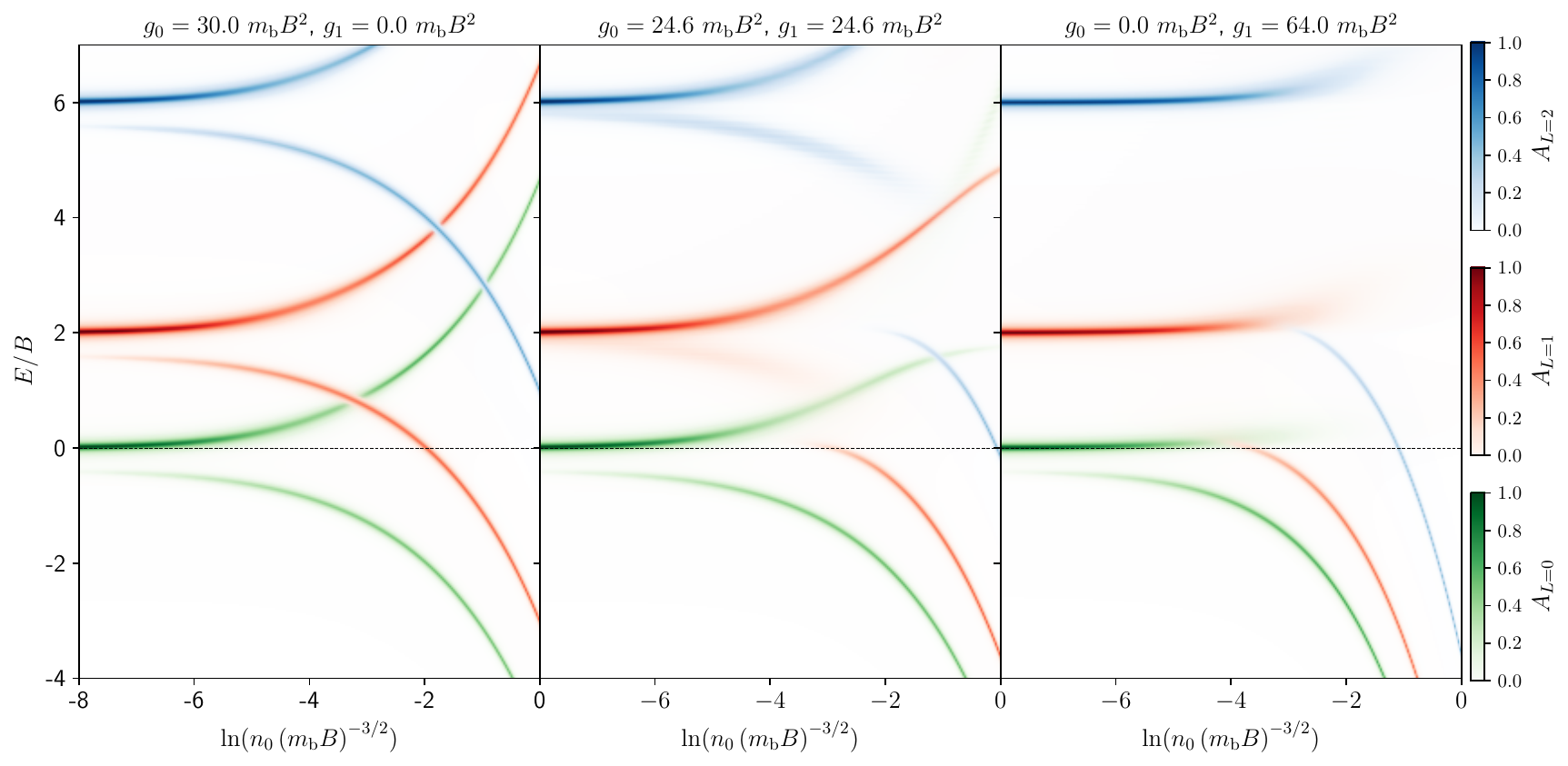}
\caption{Transition from isotropic to anisotropic interactions ($\alpha = 1.0$). Shown are spectral functions on the one-phonon level for an $s$-wave potential dominated angulon (left panel), an intermediate case with $g_0>0$ and $g_1>0$ (middle panel), and a $p$-wave potential dominated angulon (right panel). The interaction strengths (given in the titles) are chosen such that there is one bound state per $L$ block and the ground state energy at $n_0=0$ matches across the panels. The colors correspond to different $L$ blocks (up to $L=2$).}
\label{fig:s_and_p}
\end{figure*}

\subsection{Angulon picture}
We corroborate our analysis by switching to the angulon picture. In Fig. \ref{fig:spec_lin_angulon}, we show spectral features of the $p$-wave dominated angulon. The long-lived angulon states correspond to the sharp features, which reproduce the spectrum $\alpha BL(L+1)$ of a free rotor at low bath density $n_0$. We choose $g_1=50 \ m_{\mathrm{b}}B^2$ to ensure that there is exactly one bound state for $L=0$ in the low-density limit $n_0 = 0$. \par 

Once the medium becomes denser, a rich many-body structure emerges. The contrast to the $s$-wave dominated repulsive one-phonon angulon modeled using a variational approach\cite{lemeshko_schmidt_2015} is stark: We observe that the long-lived attractive angulon living in the $L=1$ block has a vanishing spectral weight in most of the $0 \leq E/B \leq 2$ window. This is a manifestation of the many-body instability which suppresses the quasiparticle weight of the bound state, see regime \ref{item_mb_instab} of Fig.~\ref{fig:hybrid_formalism}. \par 

Note that the features of the bound state calculated using a one-phonon ED are qualitatively different from those from the two-phonon ED. This is most evident for the $L=2$ block. At the one-phonon level, rotational states of the molecule are directly coupled only if they differ by one quantum of angular momentum. Consequently, the $L=2$-bound-state branch appears right below the $L=1$ continuum branch. At the two-phonon level, molecular rotational states differing by two quanta of angular momentum can be directly coupled, allowing for the formation of a much wider (in terms of energy) instability regime. We report similar findings in the polaron picture, see Appendix~\ref{sec:chevy_vs_non_chevy}.\par

For $L=1$, the ED calculations with two phonons show that the long-lived bound states are energetically much closer to the ground state ($L=0$ block) than suggested by the naive $\Delta E = 2$ estimate valid for isotropic interactions (cf. Fig.~\ref{fig_lin_vs_quasi_scatt}). The energy separation between bound states of adjacent $L$ blocks for two-phonon ED is smaller than for one-phonon ED (due to the coupling between molecular rotational states that differ by two quanta of angular momentum). The spectral weights that correspond to these bound states are likewise smaller. For calculations with larger number of phonons, we expect that the moment of inertia of a bound state 
will become larger and the energies of the bound states for $L\neq0$ will be even closer to the ground-state energy. One must exercise care however when analyzing this situation, as we work with $a_{\text{bb}}=0$, which implies that all bosons excited from the condensate can in principle be in the bound state (similar to studies of impurities in a Bose gas in Ref.~\onlinecite{Drescher2021}).

\section{Transition from Isotropic to Anisotropic Potentials}
\label{sec:s_and_p}

After discussing the limiting cases of pure $s$- and $p$-wave interactions, let us touch upon the transition between the two. For convenience, we focus on the angulon picture and one-phonon ED calculations. The left and right panels of Fig.~\ref{fig:s_and_p} present the spectral function of the two limiting cases. The middle panel is for a system where the microscopic interaction strengths $g_0$ and $g_1$ are of equal importance. Note that the values of $g_0$ and $g_1$ in the middle panel are chosen rather arbitrarily. Other values could have been used as well to illustrate a smooth transition between the left and right panels of Fig.~\ref{fig:s_and_p}.
\par 

For isotropic interactions ($g_1=0$), each $L$ block features two long-lived states. The upper one is in the continuum part of the spectrum. The other one originates from a rotor-boson bound state at $n_0=0$. The energies of the former (latter) states are increasing (decreasing) functions of the bath density $n_0$. The blocks are not coupled and the branches can freely cross one another. This changes if $g_1\neq 0$ in which case the bound state branch for $L>0$ blocks has an instability regime driven by an avoided crossing of energy levels. The bound states for $L>0$ reappear below the continuum branch of the $(L-1)$ block. However, this is an artifact of the one-phonon ansatz. If one used a two-phonon ansatz, the $L=2$ bound state would feature an additional avoided crossing with the $L=0$ block, reappearing below its continuum branch, see Fig.~\ref{fig:spec_lin_angulon}.\par 

In the limiting case of purely anisotropic interactions ($g_0=0$), the discontinuity becomes maximal, and the $L>0$ bound state branches vanish for intermediate and small bath densities (see also Sec.~\ref{sec:p_wave}). In other words, when the $L$ block bound state is inside the $(L-1)$ block continuum branch, the system enters an instability for sufficiently strong anisotropic interactions. A coupling between molecular rotational states differing by two quanta of angular momentum can be constructed either by adding more phonons or by adding higher-order multipoles to the interaction potential~\eqref{V_model_pot}.

\section{Summary and Outlook}
\label{sec:concl}

We studied a rotating impurity in a Bose gas assuming that the impurity-boson potential can support a bound state. We developed a beyond-linear-coupling model, and analyzed it using a numerical method based upon exact diagonalization with one and two phonons.

First, we focused on pure $s$-wave interactions. In this case, the angulon problem is equivalent to a static impurity in a Bose gas, connecting our problem to the body of known polaron results. Secondly, we analyzed pure $p$-wave interactions. We demonstrated that the many-body instability can destroy the angulon quasiparticle if there is a shallow molecule-boson bound state.\par 

Our results are based on the assumption that $a_{\text{bb}}=0$, which leads to a fast convergence of our ED method, and allows for the most direct illustration of our results. It is natural to expect that strong angular momentum transfer facilitated by the presence of a shallow bound state will be present also for $a_{\text{bb}}\neq 0$. Still, further investigations of the effect of $a_{\text{bb}}$ are needed to identify the most realistic experimental scenario for confirming our findings.
Another assumption of our work is that there is a single bound state in the system. However, for heavy molecules immersed in helium droplets\cite{Barnett1994}, a shallow bound state can actually be an excited state of the molecule-helium-atom system. Further work is needed to understand this scenario, in particular, the effect of deep bound states on our results. 
\par

Finally, we want to outline a few other research directions motivated by our study. At the two-body level (one boson and a molecule), it appears interesting to develop an effective description of the system in the vicinity of the threshold for binding, which can be later used to parametrize many-body results. This problem can be approached in the spirit of  effective field theories (EFT)~\cite{Hammer2020}, which establish correlations between different observables, such as binding energies and low-energy scattering parameters. EFT can also help to understand three-body physics (for two bosons and a molecule), which might be relevant for light molecules where active translational degrees of freedom pave the way for the Efimov effect~\cite{EFIMOV1970563,BRAATEN2006259,NIELSEN2001373}.\par 

The few-body physics becomes very different if we restrict rotational and translation motions to a two-dimensional geometry, where any $p$-wave potential supports a bound state~\cite{SIMON1976279,Volosniev2011a,Volosniev2011}. The presence of this bound state may modify the known properties of a planar angulon~\cite{Yakaboylu2018}, motivating further studies of $p$-wave interacting two-dimensional rotors. Finally, it might be interesting to analyze the effect of shallow bound states on the time dynamics that can be realized in a laboratory~\cite{lemeshko_2013,Koch2019}. It is known that the dynamics of the Bose polaron is strongly affected by the presence of a bound state~\cite{Volosniev2015,Shchadilova2016,Drescher2021} (compare also Refs.~\onlinecite{Skou2021} and~\onlinecite{Morgen2023}), which implies that time evolution of the angulon quasiparticle may contain information about shallow bound states observable in a laboratory.

\section*{Data Availability}
High-level data products are made available on reasonable request. For implementation details please consult the Cython implementation on the public GitHub repository \href{https://github.com/tibordome/ed-angulon}{https://github.com/tibordome/ed-angulon}.

\section*{Acknowledgements}
We would like to thank Giacomo Bighin, Igor Cherepanov, Ekaterina Paerschke and Enderalp Yakaboylu for insightful discussions on a wide range of topics. This work has been supported by the European Research Council (ERC) Starting Grant No. 801770 (ANGULON). A.G. and A.~G.~V.~acknowledge support from the European Union's Horizon 2020 research and innovation program under the Marie Sk\l{}odowska-Curie grant agreement No. 754411. Numerical calculations were performed on the Euler cluster managed by the HPC team at ETH Zurich. R.S. acknowledges support by the Deutsche Forschungsgemeinschaft under Germany's
Excellence Strategy EXC 2181/1-390900948 (the Heidelberg STRUCTURES Excellence Cluster). T.~D. acknowledges support from the Isaac Newton Studentship and the Science and Technology Facilities Council under grant number ST/V50659X/1.    

\appendix

\section{One-Phonon vs Two-Phonon Results}
\label{sec:chevy_vs_non_chevy}

\begin{figure}[t]
\begin{minipage}{0.5\textwidth}
\includegraphics[scale=.45]{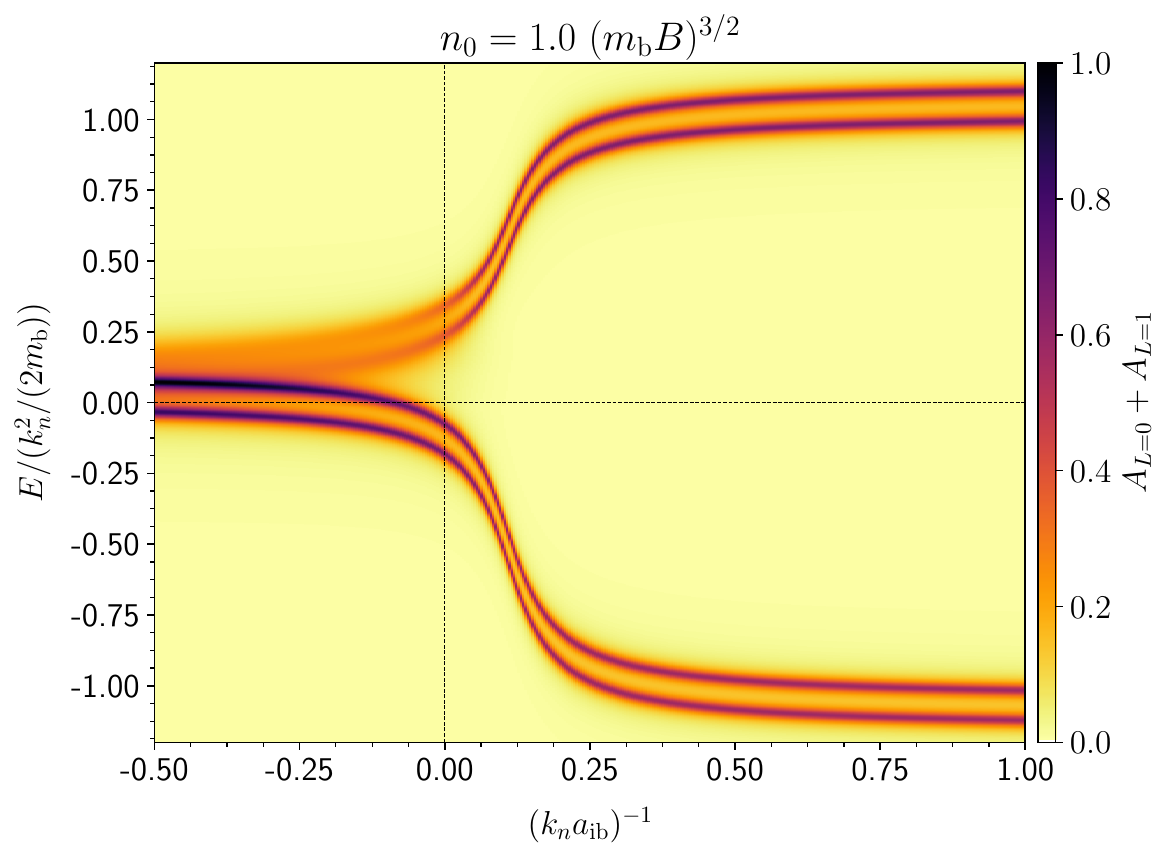}
\end{minipage}
\begin{minipage}{0.465\textwidth}
\includegraphics[scale=.45]{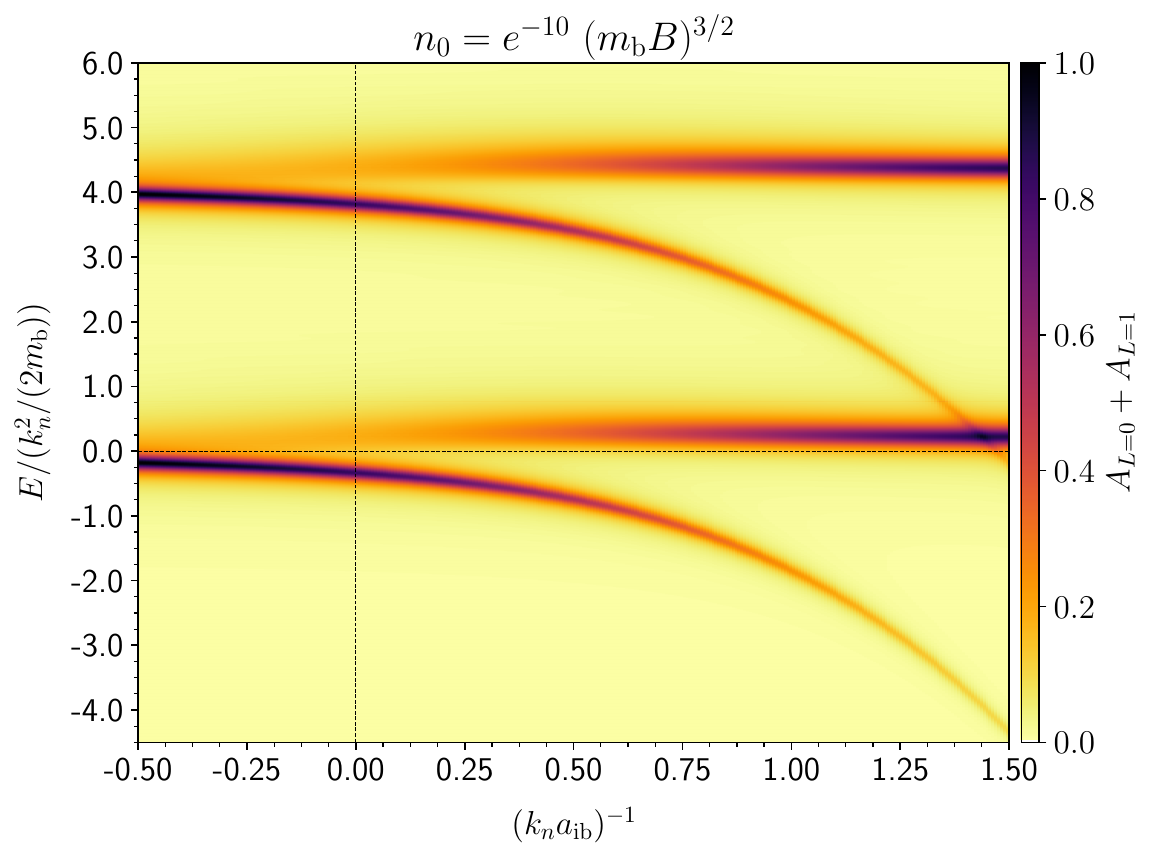}
\end{minipage}
\caption{Spectral function of the $s$-wave dominated system ($g_1=0$) calculated with one-phonon ED for the $L=0,1$ blocks. Top panel: $n_0 = 1.0 \ (m_{\mathrm{b}}B)^{3/2}$ and $\alpha = 0.4$.
Bottom panel: $n_0 = e^{-10} \ (m_{\mathrm{b}}B)^{3/2}$ and $\alpha = 0.02$.
Note that the repulsive polaron is visible for low bath densities but morphs into a fork-like structure when increasing $n_0$.}
\label{fig:N1_polaron}
\end{figure}

In this appendix, we show the differences between a one- and two-phonon ED approach and discuss finite-range effects. In Fig.~\ref{fig:N1_polaron} (top panel), we reduce the maximum number of phonon excitations to one while keeping all other numerical and microscopic parameters in line with the two-phonon results in Fig.~\ref{fig_lin_vs_quasi_scatt}. As in Sec.~\ref{sec:s_wave}, the $L=0,1$ blocks exhibit identical spectral features modulo an overall offset in energy due to the rotational kinetic energy in the $L=1$ block.\par 

In the figure, we also present results for a dilute gas with $n_0 = e^{-10} \ (m_{\mathrm{b}}B)^{3/2}$ [see the bottom panel of Fig.~\ref{fig:N1_polaron} and cf. to a much denser gas with $n_0 = 1.0 \ (m_{\mathrm{b}}B)^{3/2}$ in the top panel of Fig.~\ref{fig:N1_polaron}]. Independent of the density, we identify a long-lived ``attractive polaron'' feature in each $L$ block which flattens off at the values of the rescaled scattering length $(k_n a_{\text{ib}})^{-1}$ for which finite-range effects become important. Comparing the top panel of Fig.~\ref{fig:N1_polaron} to the two-phonon case presented in Fig.~\ref{fig_lin_vs_quasi_scatt}, we notice that the `finite-range' plateau occurs at higher energies at the one-phonon level. This is expected as two phonons must lead to a deeper bound state if $a_{\mathrm{bb}}=0$.\par

Note that in the limit of a dilute Bose impurity system (bottom panel of Fig.~\ref{fig:N1_polaron}), the spectral function features a well-known ``repulsive polaron'' with a strong spectral signature away from unitarity. The repulsive polaron in a dilute Bose gas can thus be described using the one-phonon excitation ansatz in agreement with studies based on contact interactions~\cite{Rath_2013,Shchadilova2016}. However, here we show that due to the finite-range molecule-boson interaction potential employed in our study, Eq.~\eqref{V_model_pot}, some different features arise that are not present in the contact interaction model. Specifically, the ``repulsive polaron'' feature has a ``nonstandard'' structure as $n_0$ grows, see the top panel of Fig.~\ref{fig:N1_polaron}. For example, for $n_0 = 1.0 \ (m_{\mathrm{b}}B)^{3/2}$ we observe that the `attractive and repulsive polaron' features form a two-pronged fork in the spectral function landscape. While the upper prong in the phonon continuum (`repulsive polaron') has a lower spectral weight close to the unitarity limit compared to the long-lived attractive polaron, it is pronounced for smaller values of the scattering length. At the two-phonon level (cf. Fig. \ref{fig_lin_vs_quasi_scatt}), the upper prong is less noticeable due to the `repulsive polaron' feature that appears with much higher spectral weight.\par 

Let us briefly explain why a reduction of the density $n_0$ effectively leads to a decrease in the effective molecule-boson interaction range and the appearance of the standard repulsive polaron feature (see the bottom panel of Fig.~\ref{fig:N1_polaron}).
At zero temperature, the natural length scales of our problem are given by $r_0$ and $n^{-1/3}$. Therefore, we expect that
the energy $E = n^{2/3} F(g_0, n^{1/3}r_0)$ of a finite-range model can be expressed as a function $F$ that depends only on the combination $n^{1/3}r_0$. To illustrate this, we write the Schr\"odinger equation $H\Psi(\mathbf{x}) = E\Psi(\mathbf{x})$ for the boson-molecule problem
\begin{align}
\begin{aligned}
& n^{-2/3}H\Psi(\mathbf{x}) =-n^{-2/3}\left(\frac{1}{2}\frac{\partial^2}{\partial \mathbf{x}^2} + \frac{g_0}{2r_0^2}e^{-\mathbf{x}^2/r_0^2}\right)\Psi(\mathbf{x}) \\
&=-\left(\frac{1}{2}\frac{\partial^2}{\partial (n^{1/3}\mathbf{x})^2} + \frac{g_0}{2(n^{1/3}r_0)^2}e^{-\frac{(n^{1/3}\mathbf{x})^2}{(n^{1/3}r_0)^2}}\right)\Psi(\mathbf{x}) \\
&=-\left(\frac{1}{2}\frac{\partial^2}{\partial \tilde{\mathbf{x}}^2} + \frac{g_0}{2\tilde{r_0}^2}e^{-\frac{\tilde{\mathbf{x}}^2}{\tilde{r_0}^2}}\right)\tilde{\Psi}(\mathbf{\tilde{x}}),
\end{aligned}
\end{align}
where we have defined $\tilde{\mathbf{x}}=n^{1/3}\mathbf{x}$ and $\tilde{r}_0=n^{1/3}r_0$. Thus, the expectation value of the energy (and other relevant observables) up to a multiplicative factor can be expressed as $ F(g_0, n^{1/3}r_0)$. The parameter $g_0$ introduces other (effective) length scales in the problem, in particular, the scattering length; it is fixed by two-body physics independently of the density of the Bose gas. As in our investigation we work close to unitarity, all length scales except the scattering length become irrelevant: one can use $g_0\simeq f(a_{\mathrm{ib}}/r_0)$ (cf. Ref.~\onlinecite{Jeszenszki_2018}). Therefore, the reduction of the density with fixed $k_na_{\mathrm{ib}}$
corresponds to a reduced effective range. 

\begin{figure}[t]
\includegraphics[scale=.57,trim={0.2cm 0 0 0},clip]{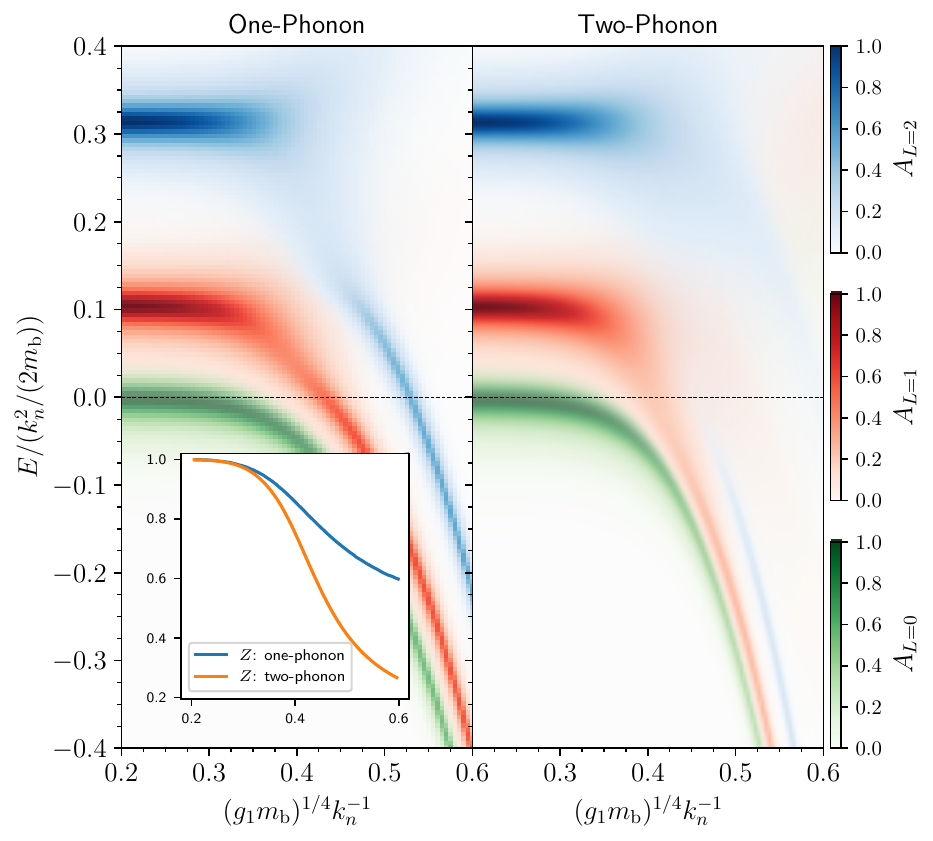}
\caption{Decomposition of the spectral function of the $p$-wave dominated  system ($g_0=0$) into the $L=0, 1, 2$ blocks (differentiated by the color maps) with $\alpha = 0.4$ and bath density $n_0 = 1.0 \ (m_{\mathrm{b}}B)^{3/2}$. Left panel: results for one-phonon ED. Right panel: results for two-phonon ED. The inset shows the quasiparticle weight $Z$ for the ground state of the $L=0$ block as a function of $(g_1m_{\mathrm{b}})^{1/4}k_n^{-1}$.}
\label{fig:N1_vs_N2_hybrid}
\end{figure}

We now investigate one-phonon vs two-phonon results in the purely anisotropic $p$-wave potential limit. The one-phonon ED partially captures the qualitative features of the ``quantum carousel'', as shown in Fig.~\ref{fig:N1_vs_N2_hybrid}. While the dressed rotor in the weak-coupling regime is well recovered, the long-lived bound state branch appears at much higher energies than in the two-phonon case. Since the single available phonon is preferably attached to the impurity if the coupling allows, there is simply no phonon `left' to properly account for the impurity-bath coupling. The moment of inertia of the bound state is thus underestimated, see Sec.~\ref{sec:p_wave}. As a consequence, the spectral discontinuity is too small and the energy separation between the bound states of the $L=0$ and $L>0$ blocks is too large.\par

Another shortcoming of the one-phonon approach is that the residue is overestimated. In the strong-coupling regime attained for large values of $g_1$, the residue $Z$ for the ground state (shown in Fig. \ref{fig:N1_vs_N2_hybrid}) converges to lower values for two-phonon ED, indicating a higher level of hybridization with the bath.\par 

The instability regime and quasiparticle weights can thus be described more reliably by a two-phonon ansatz and beyond. This further corroborates the need for light and fast analytic methods such as ED that allow for an easy implementation of higher-order phonon excitations.

\section{Hamiltonian}
\label{sec:ham_deriv}

We start from a first-principle laboratory-frame Hamiltonian that describes a single linear molecule interacting with a homogeneous BEC:
\begin{equation}
\hat{\mathcal{H}}^{\text{lab. fr.}} = \hat{H}^{\text{mol}}+\hat{H}^{\text{bos}}+\hat{H}^{\text{mb}}.
\label{lframe}
\end{equation}
In units where $\hbar \equiv 1$, the kinetic energy of a linear-rotor impurity reads $\hat{H}^{\text{mol}}=\alpha B\hat{\mathbf{J}}^2$, where $B$ is the rotational constant and $\hat{\mathbf{J}}$ is the laboratory-frame angular momentum operator. The quantum state of a rigid linear rotor is defined by the eigenvalues of $\hat{\mathbf{J}}^2$ and one of the projections onto the laboratory-frame coordinate axes, usually chosen to be $\hat{J}_Z$, such that
\begin{align}
\begin{aligned}
\hat{\mathbf{J}}^2 \ket{j, m} & = j(j+1)\ket{j, m} \\
\hat{J}_Z \ket{j, m} & = m\ket{j, m}.
\end{aligned}
\label{linear_q_num}
\end{align}
In the absence of external fields, the eigenstates of a rigid linear rotor thus form $(2j+ 1)$-fold degenerate multiplets with energies given by $E_j=Bj(j+ 1)$. Often it is convenient to work in the angular representation, where the linear rotor wavefunctions are given by spherical harmonics (see Appendix~\ref{sec:am_algebra} and also Varshalovich et al.\cite{varshalovich}),
\begin{equation}
\braket{\theta, \phi}{j,m}=Y_{jm}(\theta, \phi).
\label{Y_wavefcts}
\end{equation} 

As usual, we assume that only two-body forces between the bosonic atoms are important so that the Hamiltonian that describes bosons is written in second quantization as
\begin{equation}
\hat{H}^{\text{bos}} = \sum_{\mathbf{k}}\epsilon(\mathbf{k})\hat{a}_{\mathbf{k}}^{\dagger}\hat{a}_{\mathbf{k}}+\frac{1}{2}\sum_{\mathbf{k}, \mathbf{k}', \mathbf{q}}V_{\text{bb}}(\mathbf{q})\hat{a}_{\mathbf{k}'-\mathbf{q}}^{\dagger}\hat{a}_{\mathbf{k}+\mathbf{q}}^{\dagger}\hat{a}_{\mathbf{k}'}\hat{a}_{\mathbf{k}}.
\label{Ham_bos}
\end{equation}
One can use $\sum_{\mathbf{k}}\equiv \int_{\mathbb{R}^3} \frac{d^3\mathbf{k}}{(2\pi)^3}$ to simplify calculations. Since momentum carries units of wavenumbers, each summation in Eq.~\eqref{Ham_bos} carries a dimensionality of $\left[\text{length}\right]^{-3}$. The boson creators and annihilators satisfy the canonical commutation relations (CCRs):
\begin{align}
\begin{aligned}
\left[\hat{a}_{\mathbf{r}},\hat{a}_{\mathbf{r}'}^{\dagger}\right]&=\delta^{(3)}(\mathbf{r}-\mathbf{r}') \\
\left[\hat{a}_{\mathbf{k}},\hat{a}_{\mathbf{k}'}^{\dagger}\right]&=(2\pi)^{3}\delta^{(3)}(\mathbf{k}-\mathbf{k}').
\end{aligned}
\label{ccrs}
\end{align}
Note that, in this convention, the operators $\hat{a}_{\mathbf{k}}^{\dagger}$ and $\hat{a}_{\mathbf{k}}$ are not dimensionless, but carry a dimension of $\left[\text{length}\right]^{\frac{3}{2}}$.

The first term of Eq.~\eqref{Ham_bos} is the kinetic energy of bosons, $\epsilon(\mathbf{k}) = \epsilon(k)=k^2/(2m_{\text{b}})$, with $m_{\text{b}}$ the boson mass. The second term describes boson-boson interactions, whose strength in momentum space is approximated by a pseudo-potential $V_{\text{bb}}(\mathbf{q})=g_{\text{bb}}$. We keep it for the sake of discussion while in the main text we set $V_{\text{bb}}=0$. 
Within the first-order Born approximation of the Lippmann-Schwinger equation\cite{CREMER2012167}, we identify a simple relation with the boson-boson scattering length, $a_{\text{bb}} = g_{\text{bb}}m_{\text{b}}/(4\pi)$.\par 
To make use of the macroscopic occupation of the zero-momentum quantum state, it is convenient to split the creation and annihilation operators into the zero-momentum and finite-momentum parts by
\begin{equation}
\hat{a}_{\mathbf{k}}=(2\pi)^3\hat{\Phi}_0\delta^{(3)}(\mathbf{k})+\hat{\Phi}_{\mathbf{k}\neq 0},
\label{bog_part_1}
\end{equation} 
where the factor of $(2\pi)^3$ appears due to the adopted convention for the Fourier transform, 
\begin{equation}
\hat{f}(\mathbf{R})=\int_{\mathbb{R}^3}\frac{d^3\mathbf{k}}{(2\pi)^3}\hat{f}(\mathbf{k})e^{i\mathbf{k}\cdot \mathbf{R}}.
\label{def_FT}
\end{equation}
Let us now apply the Bogoliubov approximation to Eq.~\eqref{bog_part_1}, which first replaces the operators $\hat{\Phi}_0$ and $\hat{\Phi}_0^{\dagger}$ with the $c$ number $\sqrt{n_0}$, where $n_0$ is the classical number density of particles in the condensate:
\begin{equation}
\hat{\Phi}_0 \  \rightarrow \  \sqrt{n_0}, \ \ \ \ \hat{\Phi}_0^{\dagger}  \ \rightarrow  \ \sqrt{n_0}.
\label{n_0_repl}
\end{equation}
This is equivalent to ignoring the non-commutativity of the operators $\hat{\Phi}_0$ and $\hat{\Phi}_0^{\dagger}$, which is a good approximation for describing the macroscopic phenomena that BECs exhibit. The second step of the Bogoliubov approximation is to assume that the population of excited states is small, i.e. $\sum_{\mathbf{k}}\langle | \hat{\Phi}_{\mathbf{k}}|^2\rangle \ll n_0$. Consequently, one can neglect the terms cubic and quartic in $\hat{\Phi}_{\mathbf{k}}$. The celebrated Bogoliubov transformation of the field operators reads as
\begin{equation}
\hat{\Phi}_{\mathbf{k}} = u_k\hat{b}_{\mathbf{k}}+v_k\hat{b}_{-\mathbf{k}}^{\dagger}
\label{Bog_trafo}
\end{equation}
with the real coefficients $u_k$ and $v_k$ that obey the normalization condition $| u_k |^2 - | v_k| ^2 = 1$ on the basis of the CCRs \eqref{ccrs}. This transformation diagonalizes the leading terms in Eq.~\eqref{Ham_bos},
\begin{equation}
\hat{H}^{\text{bos}} \simeq  \sum_{\mathbf{k}\neq 0}\omega(k)\hat{b}_{\mathbf{k}}^{\dagger} \hat{b}_{\mathbf{k}}.
\label{Ham_bos_diag}
\end{equation}
Note that Hamiltonian \eqref{Ham_bos_diag} provides a useful description only for weak impurity-boson interactions. It is also not useful for some other scenarios,  e.g., in the presence of vortices \cite{Dulieu_2017}.

The Bogoliubov coefficients read as
\begin{align}
\begin{aligned}
u_k&=\left(\frac{\epsilon (k)+V_{\text{bb}}(k)n_0}{2\omega(k)}+\frac{1}{2}\right)^{\frac{1}{2}}, \\
v_k&=-\left(\frac{\epsilon (k)+V_{\text{bb}}(k)n_0}{2\omega(k)}-\frac{1}{2}\right)^{\frac{1}{2}}.
\end{aligned}
\label{Bog_solved}
\end{align}
Equation~\eqref{Ham_bos_diag} describes the bosonic system in terms of non-interacting Bogoliubov quasiparticles (or Bogoliubons) with the dispersion relation 
\begin{equation}
\omega(k)= \sqrt{\epsilon(k)\left[\epsilon(k)+\frac{8\pi a_{\text{bb}}n_0}{m_{\text{b}}}\right]}.
\label{e_bog_disp}
\end{equation}
Although $\omega(k)$ is only linear for small momenta $k \ll m_{\text{b}}c = \sqrt{m_{\text{b}}g_{\text{bb}}n_0}$, we refer to both particle-like and wave-like limits of $\omega(k)$ as ``phonons''.\par
Most generally, the interaction between an impurity and the bosonic atoms is given by
\begin{equation}
\hat{H}^{\text{mb}} = \sum_{\mathbf{k}, \mathbf{q}}\hat{V}_{\text{mb}}(\mathbf{q}, \hat{\phi}, \hat{\theta}, \hat{\gamma})\hat{\rho}(\mathbf{q})\hat{a}^{\dagger}_{\mathbf{k}+\mathbf{q}}\hat{a}_{\mathbf{k}},
\label{mb}
\end{equation}
where $\hat{\rho}(\mathbf{q})=1$ is Fourier transform of the Dirac $\delta^{(3)}$ function, which describes the density of an immobile, pointlike impurity located at $\mathbf{r}=0$. The boson operators $\hat{a}_{\mathbf{k}}$ satisfy the CCRs \eqref{ccrs}. As opposed to atomic impurities in cold-atom systems, the microscopic interaction potential
\begin{equation}
V_{\text{mb}}(\mathbf{r})=\sum_{\lambda}V_{\lambda}(r)Y_{\lambda 0}(\theta_r, \phi_r)
\end{equation}
between a linear molecule and a boson, where $\mathbf{r}=(r, \theta_r, \phi_r)\equiv (r, \hat{\mathbf{r}})$ is the boson coordinate vector in the molecular frame $(x,y,z)$, permits angular momentum exchange via $V_{\lambda}$.\par 
The Wigner $D$ matrices (see Appendix~\ref{sec:am_algebra}) transform covariant components of any irreducible tensor. Combining this insight with the plane-wave expansion, the spherical harmonic addition theorem, the Bogoliubov approximation \eqref{bog_part_1} and transformation \eqref{Bog_trafo} as well as the angular momentum representation [Eq. \eqref{basis_change}] we arrive at 

\footnotesize
\begin{align}
\begin{aligned}
&\hat{H}^{\text{mb}}= \sum_{\mathclap{k\neq 0, \lambda \mu}}U_{\lambda}(k)\sqrt{\frac{4\pi}{2\lambda+1}}\left(\hat{Y}_{\lambda \mu}^{\dagger}(\hat{\Omega})\hat{b}^{\dagger}_{k\lambda\mu}+\text{H.c.}\right)\\
& \ \ +\sum_{\mathclap{\substack{k, q\neq 0\\ \lambda \mu l m l' m'}}}W_{l'\lambda}^{l}(k,q)C^{lm}_{l'm'\lambda\mu}(u_ku_q+v_kv_q)\hat{Y}_{\lambda \mu}^{\dagger}(\hat{\Omega})\hat{b}^{\dagger}_{klm}\hat{b}_{ql'm'}\\
& \ \ +\sum_{\mathclap{\substack{k, q\neq 0\\ \lambda \mu l m l' m'}}}W_{l'\lambda}^{l}(k,q)C^{lm}_{l'm'\lambda\mu}(-1)^{m'}u_kv_q\left(\hat{Y}_{\lambda \mu}^{\dagger}(\hat{\Omega})\hat{b}^{\dagger}_{klm}\hat{b}^{\dagger}_{ql'-m'}+\text{H.c.}\right).
\end{aligned}
\label{intractable}
\end{align}
\normalsize
Here, $\sum_{k\neq 0}\equiv \int_0^\infty dk$. The one-phonon coupling coefficients read as
\begin{equation}
U_{\lambda}(k)= \sqrt{\frac{2k^2n_0\epsilon(k)}{\omega(k)\pi}}\int_0^\infty drr^2V_{\lambda}(r)j_{\lambda}(kr),
\label{eq:U_lambda}
\end{equation}
and the two-phonon coupling coefficients read
\small
\begin{equation}
W_{l'\lambda}^{l}(k,q)= \frac{2}{\pi}kq\sqrt{\frac{2l'+1}{2l+1}}C^{l0}_{l'0\lambda 0}\int_{\mathbb{R}_+}dr r^2 V_{\lambda}(r)j_l(kr)j_{l'}(qr).
\end{equation}
\normalsize
The real scalars $C^{l\nu}_{l'\nu'\lambda\mu}$ are the Clebsch-Gordan coefficients, and $j_l$ are spherical Bessel functions of the first kind.\par

To perform calculations, it is convenient to exploit the conservation of the total angular momentum $\hat{\mathbf{L}} = \hat{\mathbf{J}}+\hat{\text{\boldmath$\Lambda$}}$, where $\hat{\text{\boldmath$\Lambda$}}$ is the composite angular momentum of the bath:
\begin{equation}
\hat{\text{\boldmath$\Lambda$}} = \sum_{k\lambda \mu \nu}\hat{b}^{\dagger}_{k\lambda\mu}\mathbf{J}^{(\lambda)}_{\mu \nu}\hat{b}_{k\lambda \nu}.
\label{composite_am}
\end{equation}
Here, $\mathbf{J}^{(\lambda)}$ is the three-dimensional (3D) vector of matrices fulfilling the angular momentum algebra in the representation of angular momentum $\lambda$, say $J^{(1/2)}_z = \left(\begin{smallmatrix} 1&0\\ 0&-1 \end{smallmatrix}\right)$. As one can easily check, the natural relations
\begin{align}
\begin{aligned}
& \hat{\text{\boldmath$\Lambda$}}^2\ket{k,\lambda, \mu}^{\text{coll}}=\lambda (\lambda+1)\ket{k,\lambda, \mu}^{\text{coll}} \\
& \hat{\Lambda}_z\ket{k,\lambda, \mu}^{\text{coll}}= \mu \ket{k,\lambda ,\mu}^{\text{coll}}
\end{aligned}
\end{align}
hold for \emph{collective} bath states $\ket{k,\lambda,\mu}^{\text{coll}}$. First described in 2016~\cite{lemeshko_schmidt_2016}, the angular momentum analog of the Lee-Low-Pines (LLP) transformation~\cite{LLP_1953} is then generated by the composite angular momentum of the bath via
\begin{equation}
\hat{S}=e^{-i\hat{\phi}\otimes \hat{\Lambda}_z}e^{-i\hat{\theta}\otimes \hat{\Lambda}_y}e^{-i\hat{\gamma}\otimes \hat{\Lambda}_z}.
\label{canonical}
\end{equation}
While the Euler angle operators $(\hat{\phi}, \hat{\theta}, \hat{\gamma})$ act in the Hilbert space of the rotor, $\hat{\text{\boldmath$\Lambda$}}$ acts in the Hilbert space of the bath. By virtue of the phononic operators $\hat{b}^{\dagger}_{k\lambda \mu}$ and $\hat{b}_{k\lambda \mu}$ being defined as irreducible tensor operators, or spherical tensor operators, of rank $\lambda$, the Wigner $D$ matrices determine the transformation rule for Bogoliubov modes (point \ref{ubiquity_wigner} of Appendix~\ref{sec:am_algebra}) via
\begin{equation}
\hat{S}^{-1}\hat{b}^{\dagger}_{k\lambda \mu}\hat{S}=\sum_{\nu}\hat{D}_{\mu \nu}^{\lambda \dagger}(\hat{\Omega})\hat{b}^{\dagger}_{k\lambda \nu}.
\label{b_trafo}
\end{equation}
Combining the commutation relations \eqref{Wigner_D_J_comm} with the identity 
\begin{equation}
\sum_{m m' \mu}C^{lm}_{l'm'\lambda \mu} \hat{D}^{\lambda}_{\mu 0}\hat{D}^{l\dagger}_{m\nu}\hat{D}^{l'}_{m'\nu'}=C^{l\nu}_{l'\nu'\lambda 0},
\end{equation}
the molecular-frame Hamiltonian is found to be
\small
\begin{align}
\begin{aligned}
&\hat{\mathcal{H}}^{\text{m. fr.}}= \hat{S}^{-1}\hat{\mathcal{H}}^{\text{lab. fr.}}\hat{S}= \alpha B(\hat{\mathbf{J}}'-\hat{\text{\boldmath$\Lambda$}})^2+\sum_{\mathclap{k\neq 0,\lambda \mu}}\omega(k)\hat{b}_{k\lambda\mu}^{\dagger} \hat{b}_{k\lambda\mu}\\
&+\sum_{\mathclap{k\neq 0, \lambda}}U_{\lambda}(k)\left[\hat{b}_{k\lambda 0}^{\dagger} + \text{H.c.}\right] +\sum_{\mathclap{\substack{k, q \neq 0 \\ \lambda l l' \nu}}} {}_1W^l_{l' \lambda}(k,q)C^{l \nu}_{l'\nu \lambda 0}\hat{b}_{kl\nu}^{\dagger} \hat{b}_{ql'\nu}\\ &+\sum_{\mathclap{\substack{k, q \neq 0 \\  \lambda l l' \nu}}} {}_2W^l_{l' \lambda}(q,k)C^{l \nu}_{l'\nu \lambda 0}\left[\hat{b}_{ql\nu}^{\dagger} \hat{b}^{\dagger}_{kl'-\nu}+\text{H.c.}\right].
\end{aligned}
\label{e_mol_frame_hamiltonian}
\end{align}
\normalsize
We have introduced the effective potentials
\begin{equation}
{}_1W^l_{l' \lambda}(k,q)= (u_ku_q+v_kv_q)\sqrt{\frac{2l+1}{4\pi}}W_{l' \lambda}^l(k,q)
\end{equation}
and
\begin{equation}
{}_2W^l_{l' \lambda}(k,q) = u_kv_q\sqrt{\frac{2l+1}{4\pi}}W_{l' \lambda}^l(k,q).
\end{equation}
The projection of the laboratory-frame angular momentum operator $\hat{\mathbf{J}}$ onto the molecular-frame axes is denoted by $\hat{\mathbf{J}}'$. Its components $\hat{J}'_k$ ($k\in \lbrace -1, 0, +1 \rbrace$) can be expressed in terms of the covariant spherical basis
\begin{align}
\begin{aligned}
\mathbf{e}_{-1} &= \frac{1}{\sqrt{2}}\left(\mathbf{e}_x-i\mathbf{e}_y\right), \\
\mathbf{e}_0 &= \mathbf{e}_z, \\
\mathbf{e}_{+1} &= -\frac{1}{\sqrt{2}}\left(\mathbf{e}_x+i\mathbf{e}_y\right). \\\end{aligned}
\end{align}
The first advantage of Hamiltonian \eqref{e_mol_frame_hamiltonian} over \eqref{lframe} is that the former does not contain the impurity's Euler coordinates $(\theta, \phi)$, allowing us to bypass the intractable angular momentum algebra arising from the impurity-bath coupling, manifest in Eq.~\eqref{intractable}.\par 

Secondly, the laboratory-frame eigenstate $\ket{L,M}=\sum_{\substack{jm \\ k\lambda \mu i}}a^i_{k\lambda j}C^{LM}_{jm\lambda \mu}\ket{jm0}\otimes \ket{k,\lambda,\mu}^{\text{coll}}_i$ of total angular momentum $L$ and $Z$ projection $M$ (here the index $i$ labels a possible phonon configuration that results in the collective bath state $\ket{k,\lambda,\mu}^{\text{coll}}$) is transformed into the many-body state 
\begin{equation}
\hat{S}^{-1}\ket{L,M}=\sum_{k\lambda ni}f^i_{k\lambda n}\ket{L,M,n}\otimes \ket{k,\lambda, n}^{\text{coll}}_i,
\label{transformed_state}
\end{equation}
where the coefficients are given by $f^i_{k\lambda n}=(-1)^{\lambda+n}\sum_ja^i_{k\lambda j}C^{j0}_{L-n\lambda n}$. Each state $\ket{LMn}$ in the superposition \eqref{transformed_state} is an \emph{effective} symmetric-top state, with the projection $n$ of the total angular momentum onto the $z$ axis being entirely determined by the boson field; the canonical transformation \eqref{canonical} converts a linear-rotor molecule into an effective symmetric top by dressing it with a boson field. Hence, the Hamiltonian \eqref{full_mol} is explicitly expressed through the total angular momentum, a constant of motion, evident in $\hat{\mathbf{J}}^{'2}\hat{S}^{-1}\ket{L,M}=L(L+1)\hat{S}^{-1}\ket{L,M}$.\par

Thirdly, in the strong-coupling limit of $\xi = B/\max\limits_{k, \lambda}U_{\lambda}(k)\ll 1$, the Hamiltonian \eqref{full_mol} can be exactly solved by means of an additional coherent state transformation~\cite{lemeshko_schmidt_2016}, provided quadratic terms are excluded. As a comparison, the beyond-Fr\"ohlich static Bose polaron can be solved exactly in one dimension~\cite{Kain_2018} and the ground polaron state is found to be an exact multimode squeezing state. However, in the case of the Bose angulon, the saddle point equation for the mean-field variational parameters cannot be solved in closed form, prohibiting an exact solution of the beyond-Fr\"ohlich static Bose angulon even in the limit of $\xi \ll 1$. Exact diagonalization (ED) is thus one of the few ways to make progress.\par 
In the Lehmann representation (which corresponds to an expansion in the many-particle eigenstates under the assumption of a time-independent Hamiltonian~\cite{mahan}), the retarded Green's function of the angulon reads as
\begin{equation}
G^{\text{ret,ang}}(E)=\lim_{\epsilon \rightarrow 0^+} \ \sum_L \sum_{j}\frac{|\braket{\Psi_{\mathrm{NI}}}{\Psi_{L}^{(j)}}|^2}{E-\varepsilon_L^{(j)}+i\epsilon}.
\label{comp_greens_function}
\end{equation}
Here, the $j$ index runs over all eigenstates $\Psi_L^{(j)}$ with associated eigenvalues $\varepsilon_L^{(j)}$ of $\hat{\mathcal{H}}^{\text{m. fr.}}$ within the $L$ block, and in order to make the set of eigenstates \emph{complete}, we need to sum over the quantum number $L$ as well. Recall that the non-interacting state $\Psi_{\mathrm{NI}}$ is the one containing no phonons. The zero-momentum angulon spectral function then reads as~\cite{Rath_2013}
\begin{equation}
A(E) = \ -\frac{1}{\pi} \text{Im} \ G^{\text{ret,ang}}(E)
\end{equation}
and obeys the sum rule $\int dE A(E) =1$, where the integration extends over all frequencies. To prove the sum rule, note that
\begin{equation}
\text{Im}\lim_{\epsilon \rightarrow 0^+}\frac{1}{x+i\epsilon}=-\pi\delta(x).
\end{equation}

\section{Angular Momentum Representation and Phonon Density Profiles}
\label{sec:am_representation}

Phonon creation and annihilation operators, $\hat{b}^{\dagger}_{\mathbf{k}}$ and $\hat{b}_{\mathbf{k}}$, can be mapped from Cartesian coordinates $\mathbf{k}=\lbrace k_x,k_y,k_z\rbrace$ to spherical coordinates $\mathbf{k}=\lbrace k, \theta_k, \phi_k\rbrace$. Yet, for the problem at hand, we are interested in the angular momentum properties of a rotating impurity immersed in a condensate. Hence, it is much more convenient to work in the angular momentum representation for the single-particle basis instead of the Cartesian or spherical one. As we will now derive, the single-particle basis change
\begin{align}
\begin{aligned}
\hat{b}_{k\lambda \mu}^{\dagger}&=\frac{ki^{-\lambda}}{(2\pi)^{\frac{3}{2}}}\int_{S^2}d\phi_kd\theta_k\sin\theta_kY_{\lambda\mu}(\theta_k, \phi_k)\hat{b}_{\mathbf{k}}^{\dagger},  \\
\hat{b}_{\mathbf{k}}^{\dagger}&=\frac{(2\pi)^{\frac{3}{2}}}{k}\sum_{\lambda \mu}i^{\lambda}Y^{*}_{\lambda \mu}(\theta_k, \phi_k)\hat{b}^{\dagger}_{k\lambda\mu},
\end{aligned}
\label{basis_change}
\end{align}
is compatible with the frequently adopted\cite{greiner} CCRs:
\begin{align}
\begin{aligned}
\left[\hat{b}_{\mathbf{k}}, \hat{b}_{\mathbf{k}'}^{\dagger}\right]& =(2\pi)^3\delta^{(3)}(\mathbf{k}-\mathbf{k}'),\\
\left[\hat{b}_{k\lambda \mu}, \hat{b}^{\dagger}_{k'\lambda'\mu'}\right]&=\delta(k-k')\delta_{\lambda \lambda'}\delta_{\mu \mu'}.
\end{aligned}
\label{comm_rels}
\end{align}
The presence of $k$'s in the transformation rule \eqref{basis_change} can be made plausible via $\delta^{(3)}(\mathbf{r}-\mathbf{r}')=\frac{1}{r^2}\delta(\theta-\theta')\delta(\phi-\phi')\delta(r-r')$. Although the transformation does not depend on the radial basis functions, it is sensitive to the chosen angular basis functions. The results above are obtained for the wave functions
\begin{equation}
\psi_{k\lambda \mu}(r, \theta_k,  \phi_k)=\sqrt{\frac{2}{\pi}}kj_{\lambda}(kr)Y_{\lambda \mu}(\theta_k, \phi_k).
\end{equation}
One possibility to derive Eq.~(\ref{basis_change}) is to consider the action of the operators $\hat{b}^{\dagger}_{\mathbf{k}}$ and $\hat{b}_{\mathbf{k}}$ on the vacuum $\ket{0}$. We ignore the fact that the eigenstates of the momentum operator are non-normalizable states,
\begin{equation}
\bra{\mathbf{r}}\hat{b}_{\mathbf{k}}^{\dagger}\ket{0}= \braket{\mathbf{r}}{\psi_{\mathbf{k}}}=\frac{e^{i\mathbf{k}\mathbf{r}}}{(2\pi)^{\frac{3}{2}}}.
\end{equation}
We also know that $\hat{b}_{k\lambda \mu}^{\dagger}\ket{0}= \ket{\psi_{k\lambda\mu}}$ satisfies
\begin{equation}
\hat{b}_{k\lambda \mu}^{\dagger}\ket{0}=\sum_{\mathbf{k}'}\braket{\psi_{\mathbf{k}'}}{\psi_{k\lambda \mu}}\ket{\psi_{\mathbf{k}'}}\equiv\sum_{\mathbf{k}'}\braket{\psi_{\mathbf{k}'}}{\psi_{k\lambda \mu}}\hat{b}_{\mathbf{k}}^{\dagger}\ket{0}.
\end{equation}
By inserting the plane-wave expansion combined with the spherical harmonic addition theorem, we arrive at
\begin{align}
\begin{aligned}
\braket{\psi_{\mathbf{k}'}}{\psi_{k\lambda \mu}}&=\frac{1}{(2\pi)^{\frac{3}{2}}}\int_{\mathbb{R}^3}d^3\mathbf{r}e^{-i\mathbf{k}'\mathbf{r}}\psi_{k\lambda \mu}(\mathbf{r})\\
&=ki^{-\lambda}Y_{\lambda\mu}(\theta_{k'},\phi_{k'})\int_0^\infty drr^2j_{\lambda}(kr)j_{\lambda}(k'r).
\end{aligned}
\end{align}
The orthogonality identity $\int_0^\infty drr^2j_{\lambda}(kr)j_{\lambda'}(k'r)=\frac{\pi}{2k^2}\delta(k-k')\delta_{\lambda \lambda'}$ for the spherical Bessel functions then yields
\small
\begin{equation}
\hat{b}_{k\lambda \mu}^{\dagger}= \sum_{\mathbf{k}'}\braket{\psi_{\mathbf{k}'}}{\psi_{k\lambda \mu}}\hat{b}_{\mathbf{k}'}^{\dagger} = ki^{-\lambda}\! \! \int_{S^2}\! \! d\phi_kd\theta_k\sin\theta_kY_{\lambda \mu}(\theta_k, \phi_k)\hat{b}_{\mathbf{k}}^{\dagger}.
\end{equation}
\normalsize
In view of our CCRs \eqref{comm_rels}, we need to add a small correction to eliminate the factor $(2\pi)^3$:
\begin{equation}
\hat{b}_{k\lambda \mu}^{\dagger}=\frac{ki^{-\lambda}}{(2\pi)^{\frac{3}{2}}}\int_{S^2}d\phi_kd\theta_k\sin\theta_kY_{\lambda\mu}(\theta_k, \phi_k)\hat{b}_{\mathbf{k}}^{\dagger}.
\end{equation}
The derivation of the other expression in \eqref{basis_change} follows the same logic and will be omitted. \par
The real-space variant of the second equation in \eqref{basis_change} yields the following expression for the phonon density $n_{\text{ph}}(\mathbf{r})= \langle \hat{b}^{\dagger}_{\mathbf{r}}\hat{b}_{\mathbf{r}}\rangle$ in the impurity frame:
\begin{equation}
n_{\text{ph}}(\mathbf{r})=\sum_{\substack{\lambda \mu \\ \lambda' \mu'}}\frac{i^{-\lambda + \lambda'}}{r^2}Y_{\lambda \mu}(\theta_r, \phi_r)Y^*_{\lambda' \mu'}(\theta_r, \phi_r)\langle \hat{b}_{r\lambda \mu}^{\dagger}\hat{b}_{r\lambda'\mu'}\rangle.
\label{phon_dens_expr}
\end{equation}
Using the real-space analog of the first equation in \eqref{basis_change}, inserting a Fourier transform in the integrand and discretizing momentum, we evaluate the partial-wave contributions to
\small
\begin{equation}
\langle \hat{b}_{r\lambda \mu}^{\dagger}\hat{b}_{r\lambda'\mu'}\rangle=\frac{2i^{\lambda - \lambda'}r^2\Delta k}{\pi} \sum_{\mathclap{k,k'=k_{\text{min}}}}^{k_{\text{max}}} k k' j_{\lambda}(kr)j_{\lambda'}(k'r)\langle \hat{b}_{k\lambda \mu}^{\dagger}\hat{b}_{k'\lambda'\mu'}\rangle.
\label{rlmu_discr}
\end{equation}
\normalsize
Our ED algorithm works in the angular momentum representation as is evident from Hamiltonian \eqref{full_mol}. The expectation values $\langle \hat{b}_{k\lambda \mu}^{\dagger}\hat{b}_{k'\lambda'\mu'}\rangle$ can thus be calculated at once for the numerical ground state. At first glance, the phonon density \eqref{phon_dens_expr} is not necessarily real as the product of spherical harmonics $Y_{\lambda \mu}Y^*_{\lambda' \mu'}$ contains the complex phase $e^{i\phi_r(\mu-\mu')}$. However, the \textit{sum} in Eq.~\eqref{phon_dens_expr} is still real. To prove it, we consider
 the antiunitary time-reversal operator $\hat{\mathcal{T}}$ defined by $\hat{\mathcal{T}}\left(\alpha \ket{k,\lambda,\mu}\right)=\alpha^{\ast}\ket{k,\lambda,-\mu}$ on the single-particle level. The time-reversal operator is  an involution (as expected for any system of bosons): $\hat{\mathcal{T}}^2=\mathds{1}$. Time-reversal symmetry is also a symmetry of the system: $[\hat{\mathcal{T}}, \hat{\mathcal{H}}^{\text{m. fr.}}]=0$. Indeed, the $\hat{b}^{\dagger}\hat{b}$ term does not couple phonons with different angular momentum projections and the $\hat{b}^{\dagger}\hat{b}^{\dagger}$ term excites two phonons with opposite angular momentum projections. Using these insights one can show that
\begin{equation}
\langle \hat{b}_{k\lambda \mu}^{\dagger}\hat{b}_{k'\lambda'\mu'}\rangle = 0, \quad \text{for} \quad \mu \neq \mu',
\label{to_prove_1}
\end{equation}
and
\begin{equation}
\langle \hat{b}_{k\lambda \mu}^{\dagger}\hat{b}_{k'\lambda'\mu}\rangle = \langle \hat{b}_{k\lambda -\mu}^{\dagger}\hat{b}_{k'\lambda'-\mu}\rangle \quad \forall \ \ k,k',\lambda, \lambda', \mu.
\label{to_prove_2}
\end{equation}
Real-valuedness of $n_{\text{ph}}$ now follows from standard properties of spherical harmonics:
\begin{align}
\begin{aligned}
&Y_{\lambda \mu}Y^*_{\lambda' \mu} + Y_{\lambda -\mu}Y^*_{\lambda' -\mu}=Y_{\lambda \mu}Y^*_{\lambda' \mu} + Y_{\lambda \mu}^*Y_{\lambda' \mu}(-1)^{\mu+\mu} \\
&= 2\cdot \text{Re} (Y_{\lambda \mu}Y^*_{\lambda' \mu}) \in \mathbb{R}.
\end{aligned}
\end{align}

\section{Angular Momentum Algebra}
\label{sec:am_algebra}

Let us summarize some key results for symmetric tops, proofs of which are either straightforward or can be found in textbooks on angular momentum\cite{varshalovich}. In this appendix, the angular momentum operators $\hat{J}_i$ and $\hat{J}_i'$ of the main text turn into differential operators $\mathcal{J}_i$ and $\mathcal{P}_i$, respectively, in the Euler angle representation, i.e., on the Hilbert space of Wigner $D$ matrices. Although derived in the latter representation, the following angular momentum properties are in fact independent of the representation:
\begin{enumerate}[wide, labelwidth=!, labelindent=0pt]
\item 
\label{part_1}
The Wigner $D$ matrices $D^{j}(\mathbf{e})$ can be defined via their matrix elements as
\begin{equation}
\bra{j,m}\hat{U}(\hat{\mathbf{e}})\ket{j'm'}= \delta_{jj'}D_{mm'}^j(\mathbf{e}),
\label{wigner_matrix_elements}
\end{equation}
where the Euler angle operators $\hat{\mathbf{e}}= (\hat{\phi}, \hat{\theta}, \hat{\gamma})$ specify the rotation $\hat{U}(\hat{\mathbf{e}})=e^{-i\hat{\phi} \hat{J}_z}e^{-i\hat{\theta}\hat{J}_y}e^{-i\hat{\gamma}\hat{J}_z}$. The conjugated Wigner $D$ matrices $D^{j*}(\mathbf{e})$ are simultaneous eigenfunctions of the three mutually commuting differential operators $\boldsymbol{\mathcal{J}}^2, \mathcal{J}_z$, and $\mathcal{P}_z$:
\begin{align}
\begin{aligned}
\boldsymbol{\mathcal{J}}^2D^{j*}_{m'm}(\mathbf{e})&=jD^{j*}_{m'm}(\mathbf{e}), \\
\mathcal{J}_zD^{j*}_{m'm}(\mathbf{e})&=m'D^{j*}_{m'm}(\mathbf{e}), \\
\mathcal{P}_zD^{j*}_{m'm}(\mathbf{e})&=mD^{j*}_{m'm}(\mathbf{e}).
\end{aligned}
\label{point_1_eqs}
\end{align}
The operator $\boldsymbol{\mathcal{P}}$ describes the internal angular momentum of the rigid rotor in the body-fixed frame, obtained by projecting the physical angular momentum operator $\boldsymbol{\mathcal{J}}$ onto the body-fixed axes $\hat{\mathbf{f}}_i$.\par 
The Wigner $D$ functions thus represent wave functions of a rigid symmetric top: They are eigenfunctions of the three operators $\boldsymbol{\mathcal{J}}^2$, $\mathcal{J}_z$, and $\mathcal{P}_z$, whose eigenvalues are sufficient to describe a symmetric top with three quantum numbers, as in Eq.~\eqref{transformed_state}.
\item The action of the ladder operators $\mathcal{J}_{\pm 1}$ on the wave functions $D^{j*}_{m'm}(\mathbf{e})$ is the standard one~\cite{varshalovich}:
\begin{align}
\begin{aligned}
\mathcal{J}_{\pm 1}D^{j*}_{m'm}(\mathbf{e})&=\sqrt{j(j+1)-m(m\pm 1)}D^{j*}_{m'\pm 1m}(\mathbf{e}), \\
\mathcal{J}_zD^{j*}_{m'm}(\mathbf{e})&=m'D^{j*}_{m'm}(\mathbf{e}).
\end{aligned}
\label{diff_equs_0_1}
\end{align}
However, the internal ladder operators $\mathcal{P}_{+1}= \mathcal{P}_x + \mathcal{P}_y$ and $\mathcal{P}_{-1}= \mathcal{P}_x - \mathcal{P}_y$ act (counter-intuitively)  as step-down and step-up ``ladder'' operators, respectively:
\begin{equation}
\mathcal{P}_{\pm 1}D^{j*}_{m'm}(\mathbf{e})=\sqrt{j(j+1)-m(m\mp 1)}D^{j*}_{m'm\mp 1}(\mathbf{e}).
\label{P_counter_prop}
\end{equation}
\item The physical angular momentum operators $\mathcal{J}_i$ fulfill the standard commutation relations
\begin{equation}
\left[\mathcal{J}_i, \mathcal{J}_j\right]=i\epsilon_{ijk}\mathcal{J}_k, \ \ \ \ i,j,k \in \lbrace x,y,z\rbrace,
\end{equation}
which are equivalent to
\begin{equation}
\left[\mathcal{J}_{\mu},\mathcal{J}_{\nu}\right]=-\sqrt{2}C_{1\mu 1\nu}^{1\mu+\nu}\mathcal{J}_{\mu+\nu}, \ \ \ \ \mu, \nu \in \lbrace \pm 1, 0\rbrace.
\end{equation}
Yet, the commutation relations
\begin{equation}
\left[\mathcal{P}_i, \mathcal{P}_j\right]=-i\epsilon_{ijk}\mathcal{P}_k, \ \ \ \ i,j,k \in \lbrace x,y,z\rbrace,
\label{anomalous_1}
\end{equation}
or
\begin{equation}
\left[\mathcal{P}_{\mu},\mathcal{P}_{\nu}\right]=\sqrt{2}C_{1\mu 1\nu}^{1\mu+\nu}\mathcal{P}_{\mu+\nu}, \ \ \ \ \mu, \nu \in \lbrace \pm 1, 0\rbrace,
\label{anomalous_2}
\end{equation}
of the internal operators $\mathcal{P}_i$ qualify as \textit{anomalous}, due to the famous sign difference in the structure factors. That is, it is the operators $-\mathcal{P}_i$ that satisfy standard angular momentum commutation relations. This result is a direct consequence of the fact that the $\mathcal{J}_i$ do not commute with the rotated axes $\hat{\mathbf{f}}_i$ appearing in the definition $\mathcal{P}_i=\hat{\mathbf{f}}_i\cdot \boldsymbol{\mathcal{J}}$.
\item The operators $\mathcal{J}_i$ and $\mathcal{P}_i$ commute:
\begin{equation}
\left[\mathcal{J}_i, \mathcal{P}_j\right] = 0, \ \ \ \ \forall \ i,j \in \lbrace x,y,z\rbrace.
\end{equation}
\item The invariant squared total angular momentum operators $\boldsymbol{\mathcal{J}}^2=\boldsymbol{\mathcal{P}}^2$, i.e., the Casimir operators, satisfy
\small
\begin{equation}
\boldsymbol{\mathcal{J}}^2 = -\csc^2\theta\left(\frac{\partial^2}{\partial \phi^2}+\frac{\partial^2}{\partial \gamma^2}-2\cos \theta\frac{\partial^2}{\partial \phi \partial \gamma}\right)-\frac{\partial^2}{\partial \theta^2}-\cot \theta \frac{\partial}{\partial \theta}.
\label{j2_p2}
\end{equation}
\normalsize
\item The Wigner $D$ functions can also be defined as solutions of the differential equations
\begin{align}
\begin{aligned}
[\mathcal{J}_{\nu},D_{mm'}^j(\mathbf{e})]&= (-1)^{\nu+1}\sqrt{j(j+1)}C_{jm1-\nu}^{jm-\nu}D^j_{m-\nu m'}(\mathbf{e}), \\
[\mathcal{P}_{\nu},D_{mm'}^j(\mathbf{e})]&=\sqrt{j(j+1)}C_{jm'1\nu}^{jm'+\nu}D^j_{m m'+\nu}(\mathbf{e}),
\end{aligned}
\label{Wigner_D_J_comm}
\end{align}
where, for the sake of completeness, we note the operators at hand:
\small
\begin{align}
\begin{aligned}
\mathcal{J}_{\pm 1}&=\frac{i}{\sqrt{2}}e^{\pm i\phi}\left[\mp\cot \theta \frac{\partial}{\partial \phi}+i\frac{\partial}{\partial \theta}\pm\frac{1}{\sin \theta}\frac{\partial}{\partial \gamma}\right], \ \mathcal{J}_0=-i\frac{\partial}{\partial \phi}, \\
\mathcal{P}_{\pm 1}&=\frac{i}{\sqrt{2}}e^{\mp i\gamma}\left[\pm\cot \theta \frac{\partial}{\partial \gamma}+i\frac{\partial}{\partial \theta}\mp\frac{1}{\sin \theta}\frac{\partial}{\partial \phi}\right], \ \mathcal{P}_0=-i\frac{\partial}{\partial \gamma}.
\end{aligned}
\end{align}
\normalsize
\item 
\label{ubiquity_wigner}
The Wigner $D$ matrices appear in many studies involving angular momentum because they realize transformations of covariant components of any irreducible tensor [Eq.~\eqref{def_irr_tensor_2} itself can be taken as their definition] of rank $j$ under coordinate rotations. For instance, the spherical harmonics $\psi_{j,m}$, eigenfunctions of a quantum particle moving in a spherically-symmetric field, or of linear rotors, with angular momentum $j$ and projection $m$ transform as
\begin{equation}
\psi_{jm'}(\theta',\phi')=\sum_{m=-j}^{j}\psi_{jm}(\theta, \phi)D_{mm'}^j(\mathbf{e}),
\label{def_irr_tensor_2}
\end{equation}
where $(\theta,\phi)$ and $(\theta',\phi')$ are polar angles in the laboratory frame and body-fixed frame, respectively.
\end{enumerate}

\section{Exact Diagonalization -- Implementation Details}
\label{sec:ed_details}
Let us first motivate our adopted numerical routine. One popular choice in the literature is diagrammatic quantum Monte Carlo [DQMC]~\citep{mishchenko_2000}. It is versatile and can be applied to any dimension, any number of phonon branches, and all types of couplings. The imaginary-time Green's function is calculated with a controlled stochastic error by performing a random walk in the space of all Feynman diagrams. However, DQMC has to face the so-called \textit{sign problem}~\cite{Houcke_2010,marchand} which is typically fatal to Monte Carlo approaches: The diagrammatic expansions of the real-time Green's functions consist of complex-valued terms whose non-positive-definite weights require careful separation of signs and magnitudes when Monte Carlo sampling over them. For weak sign problems, the estimators simply converge slower, yet for strong sign problems the ratio of the statistical errors fluctuates wildly and there is little that can be done. To extract the real-time Green's function, one could instead perform \textit{analytic continuation} of imaginary-time data~\cite{Goulko_2017}, which is not straightforward and equally ill conditioned for many worthwhile cases~\cite{bergeron}. Obtaining information on excited states would also require such bad-tempered analytic continuation techniques. Another drawback of DQMC methods is that convergence can be slow.\par 

There is thus a need for lighter and faster methods such as the momentum average approximation~\cite{Covaci_2007}, or ED. ED constitutes a vastly simpler framework, circumvents the sign problem, and provides a similar level of insight into the properties of the low-lying eigenstates and eigenvalues of quantum impurity systems. It obtains Green's functions directly in real time, as opposed to imaginary time as in DQMC. Once the enumeration of basis vectors is set up, the hashing trick (explained below) allows one to efficiently represent Hamiltonian matrices or matrices corresponding to other quantities.\par

We illustrate some implementation choices for ED. An efficient representation of ket states is needed. In our implementation, we specify a ket $\ket{\psi} = \ket{\psi_{\text{imp}}} \otimes \ket{\psi_{\text{ph}}}$ via $\ket{\psi_{\text{imp}}} = \ket{L,n}$ (here $n$ is the quantum number that specifies the projection of the angular momentum onto the molecular axis) and
\begin{equation}
\ket{\psi_{\text{ph}}} = \ket{S_0,N_{S_0}, \cdots, S_{\text{max}}, N_{S_{\text{max}}}},
\label{ket_def}
\end{equation}
with $N_{S_i}$ denoting the number of phonons in state $S_i$. Note that we have neglected the good quantum number $M$, since we do not assume any external field that would break rotational symmetry. To save memory, the single-particle phonon state label $S_i\in \lbrace 0, 1, \cdots, S_{\text{max}}\rbrace $ only appears in the definition \eqref{ket_def} if it is occupied by at least one phonon. The total number of available phononic states is given by 
\begin{equation}
S_{\text{max}} +1 = \left(\frac{k_{\text{max}}-k_{\text{min}}}{\Delta k} +1\right)\sum_{l=0}^{l_{\text{max}}}(2l+1).
\end{equation}
Here $k_{\text{min}}$, $k_{\text{max}}$, $\Delta k$ and $l_{\text{max}}$ are the chosen numerical parameters for the momentum discretization and the maximal phonon angular momentum quantum number.  For given $L$ and $n$ quantum numbers, the number of \textit{a priori} conceivable phononic configurations is
\begin{equation}
\frac{(S_{\text{max}}+N_{\text{ph}})!}{S_{\text{max}}!N_{\text{ph}}!},
\end{equation}
where $N_{\text{ph}}$ is the total number of phonons. However, Eq.~\eqref{transformed_state} stipulates that the sum of the angular momentum projections of the phonons onto the molecular $z$ axis equals $n$, reducing the number of \textit{permissible} configurations substantially.\par

In an attempt to speed up the CPU-bound task of ED, we employ the ``cython.parallel'' module available in Cython. To save memory [the Hamiltonian $\hat{\mathcal{H}}^{\text{m. fr.}}$ in Eq.~\eqref{full_mol} has numerous zero entries] we put \scshape{SCIPY}\normalfont 's package ``sparse.csr\texttt{\_}matrix'' to use, which implements the compressed sparse row (CSR) format. In order to calculate eigenvalues and eigenvectors of sparse matrices, our ED algorithm calls the \scshape{SCIPY} \normalfont function ``scipy.sparse.linalg.eigsh'', which is a wrapper to ARPACK algorithms. For the angulon spectral functions in Figs.~\ref{fig_lin_vs_quasi_scatt}, \ref{fig:hybrid_formalism} and \ref{fig:spec_lin_angulon}, we compute all eigenstates $\Psi_L^{(\alpha)}$ in Eq.~\eqref{comp_greens_function} except for the one insignificant per $L$ block associated with the largest eigenvalue, avoiding two ARPACK runs per block.\par 

As an improvement to the direct access table, we resort to hashing, allowing us to use relatively small numbers as index in a table called \textit{hash table}. It is the \textit{hash function} that converts a possibly big \textit{key} from the \textit{universe} $\mathscr{U}$ into a small practical integer value from the set of \textit{hash values} $\mathscr{H}$, $h: \; \mathscr{U} \rightarrow \mathscr{H}$. A good hash function should be efficiently computable and distribute the keys uniformly over its output range. In view of our universe consisting of ket states of the form $\ket{\psi_{\text{ph}}}$ in Eq.~\eqref{ket_def}, we choose the following injective hash function:
\begin{equation}
h(\ket{\psi_{\text{ph}}})= \sum\limits_{S_i=0}^{S_{\text{max}}}(S_i+1)S_{\text{max}}^{N_{S_i}}.
\label{equ_hash}
\end{equation}
\normalsize
The sum in Eq.~\eqref{equ_hash} runs over each phonon state label $S_i$ in state $\ket{\psi_{\text{ph}}}$. Since the hash value might still become considerably large, we further save memory by retaining only the remainder of the hash values after division by some predetermined divisor~\cite{Zhang_2010}, for instance $20$.\par 

The numerical discretization parameters we adopt on the one-phonon level are $\Delta k = 0.02 \ (m_bB)^{1/2}$, $k_{\text{max}} = 4 \ (m_bB)^{1/2}$ and $\Delta k = 0.05 \ (m_bB)^{1/2}$, $k_{\text{max}} = 4 \ (m_bB)^{1/2}$ on the two-phonon level. While eigenenergies exhibit sub-percent level convergence for one-phonon ED, increasing $k_{\text{max}}$ to $k_{\text{max}} = 5 \ (m_bB)^{1/2}$ introduces at most $4$\% variations in the energy levels for two-phonon ED. We expect that similar convergence behavior holds for other quantities, including the spectral function. The infinitesimal offset along the imaginary axis when calculating the spectral function in \mbox{Eq.~\eqref{spect_func_def}} is $\epsilon = 0.05 \ B$. Our calculations are performed with $l_{\text{max}}=1$.

\bibliographystyle{apsrev4-1}
\bibliography{refs}

\end{document}